\documentclass[%
reprint,
superscriptaddress,
 amsmath,amssymb,
prl,
longbibliography,
]{revtex4-2}

\usepackage{romannum}
\usepackage{graphicx}
\usepackage{dcolumn}
\usepackage{bm}

\usepackage{siunitx}

\begin{document}

\title{\large{Diffusive buckling fronts in lattice-based metamaterials }}

\author{Jochem G. Meijer}
\email[]{jgmeijer@uchicago.edu}
\affiliation{James Franck Institute, The University of Chicago, Chicago, IL 60637}

\author{Faadil H. Shaik}
\affiliation{James Franck Institute, The University of Chicago, Chicago, IL 60637}

\author{Victoria V. McDermott}
\affiliation{James Franck Institute, The University of Chicago, Chicago, IL 60637}

\author{Heinrich M. Jaeger}
\email[]{h-jaeger@uchicago.edu}
\affiliation{James Franck Institute, The University of Chicago, Chicago, IL 60637}
\affiliation{Department of Physics, The University of Chicago, Chicago, IL 60637}

\begin{abstract}
Mechanical metamaterials can be designed to exhibit unique mechanical properties, including tunable auxetic behavior as well as multi-stability, which arise from the geometry and configuration of the constituent building blocks. 
Lattice-based metamaterials, in particular, provide lightweight platforms where local instabilities can dictate the global response, with applications in energy routing and vibration isolation. In underdamped structures, perturbations have been found to propagate as nonlinear waves, \textit{e.g.}, transition waves or solitons. Here we investigate the opposite limit of overdamped, highly dissipative lattice metamaterials. Focusing on three-dimensional structures, we uncover how buckling instabilities, triggered by compression, propagate as fronts that shape the macroscopic behavior. We demonstrate in experiments on 3D-printed simple cubic lattices how global and local buckling modes can be controlled via the lattice geometry. By incorporating viscoelastic dissipation into a 3D-continuum model, we show that strain-driven buckling fronts obey coupled reaction–diffusion equations. The diffusion and reaction coefficients, determined by local geometry, material properties, and strain, select the propagation direction and enable steering of the  fronts. This establishes a predictive and experimentally validated framework for the control of cascading mechanical instabilities in lattice-based metamaterials.
\end{abstract}

\date{\today}

\maketitle

\begin{figure*}[t!]
  \centerline{\includegraphics[width=\textwidth]{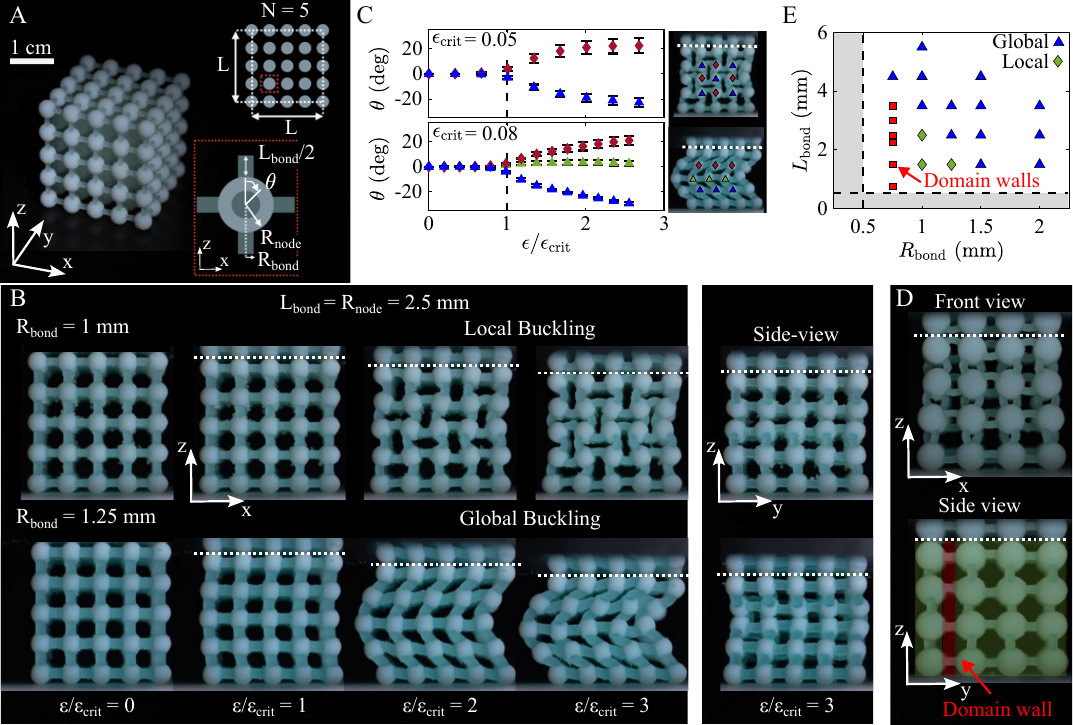}}
\caption{\textbf{3D-printed cubic lattices under compression}. 
(\textit{A}) 3D-printed cubic lattice with rigid nodes (white) and flexible bonds (lightblue) where $N = 5$, $L = \SI{30}{\mm}$, $R_{\mathrm{node}} = \SI{2.5}{\mm}$, $R_{\mathrm{bond}} = \SI{0.75}{\mm}$ and $L_{\mathrm{bond}} = \SI{2.5}{\mm}$. 
(\textit{B}) Sequence of images (front view) showing the dynamical response of two slightly different lattice structures to uniaxial compression, see Suppl.\,Movie\,1.
The overall response transitions from local to global buckling once a critical strain, $\epsilon_{\mathrm{crit}}$, is reached.
The rightmost panel shows the difference in self-folding of the five layers from the side. The white dotted lines indicate where the bottom of the compression plate touches the top of the lattice.
(\textit{C}) Node rotation $\theta$ during local and global buckling of the nine center nodes, showing counter-rotating neighbours (upper) and co-rotating rows (lower), respectively.
Clockwise rotations are defined positive (see (\textit{A})).
(\textit{D}) Front- and side view of a compressed cubic lattice with $N = 4$, $R_{\mathrm{node}} = \SI{2.63}{\mm}$, $R_{\mathrm{bond}} = \SI{0.75}{\mm}$, and $L = \SI{22.5}{\mm}$ at $\epsilon = 0.125$, see Suppl.\,Movie\,2.
Within different layers of the cube the orientation of the pattern formation changes (green regions).
This leads to torsion of the bonds (red region) connecting these layers and the formation of a domain wall.
(\textit{E}) Parameter space in terms of bond length and bond radius indicating the cubic lattices that buckle globally (blue triangles) or locally (green diamonds), including those that show the formation of a domain wall during local buckling (red squares).}
\label{fig:1}
\end{figure*}

A uniform, slender beam under sufficient compression will undergo a buckling instability, which causes a transition from a straight configuration to a new bent shape or state \cite{euler1952methodus}. 
Instead of such global deformation under load, appropriately chosen internal architectures in  mechanical metamaterials can generate instabilities that produce local deformation and/or rotation of subunits \cite{bertoldi2017flexible}.  
The coupling between the subunits  can then result in cascading instabilities that sweep through the material and transform its behavior. 
One prominent example is a plate with a dense array of circular holes in a periodic square pattern \cite{mullin2007pattern, bertoldi2008mechanics}. 
Under small uniaxial compression all holes simply flatten and become elliptical, while the overall strain response stays elastic with a positive Poisson's ratio $\nu > 0$. But beyond a critical strain, buckling leads to alternating ellipse orientations, each rotated 90 degrees with respect to its four neighbors. This generates material contraction transverse to the applied strain and thus auxetic behavior, i.e., $\nu < 0$ \cite{lakes1987foam, grima2000auxetic, grima2005auxetic,gaspar2005novel, wang2009hybrid, bertoldi2010negative,  greaves2011poisson, nicolaou2012mechanical,reid2018auxetic}.

The dynamics of cascading buckling events have so far been investigated almost exclusively in 2D structures \cite{grima2000auxetic, overvelde2012compaction,mullin2013pattern, overvelde2014relating, cho2014engineering, shan2015design, he2018buckling, cutolo2022class, dykstra2022extreme} and metabeams \cite{coulais2015discontinuous}.  
In underdamped, inertial systems it has additionally been shown that external excitations can trigger transition waves \cite{nadkarni2016unidirectional, raney2016stable, jin2020guided, yasuda2020transition, librandi2021programming} or solitons \cite{nadkarni2014dynamics, deng2017elastic, deng2018metamaterials,mo2019cnoidal, gomez2019dynamics,deng2021dynamics,deng2021nonlinear,veenstra2024non} that propagate through elastic 1D chains or 2D sytems.
These traveling perturbations can form well-characterized domain walls and can also interfere with the local architecture, allowing for the creation and tuning of bandgaps  \cite{yasuda2020transition, deng2020characterization}.
Further complexity is added by the introduction of dissipative, viscoelastic materials with inherent internal time-scales. 
This gives rise to shape-changing and snapping, controlled by the experienced strain rate \cite{nachbar1967dynamic, santer2010self, stern2018shaping, janbaz2018multimaterial,janbaz2020strain,bossart2021oligomodal,dykstra2022extreme}, leading to dissipative energy transport \cite{nadkarni2016universal} and the emergence of 1D diffusive kinks \cite{janbaz2024diffusive}.
However, so far there have been no systematic studies extending these concepts to buckling in three-dimensional lattice-based metamaterials, where such kinks generalize into strain-driven, diffusive buckling fronts.

Here, we investigate buckling in three-dimensional simple cubic lattices, fabricated by 3D-printing and subjected to uniaxial compression. 
These materials are highly dissipative and the compression occurs at a fixed rate that is sufficiently slow to mitigate the role of inertia. We map out the conditions for different buckling modes and to explain the experimental observations derive a set of coupled reaction-diffusion equations. This model, which is of a type more commonly used to analyze pattern evolution in biological or chemical systems \cite{murray1981pre,kondo2010reaction,lacalli2022patterning,mcdougal2013reaction}, dictates the emergence of local buckling events once a critical strain is reached and predicts the propagation of diffusive buckling fronts.
The diffusivity and reaction coefficients - and therefore the dynamics - depend on the local lattice geometry, material properties, and strain. 
Using this, we show how grading the properties of the lattice or introducing defects makes it possible to steer the fronts through the metamaterial, in a targeted and controllable manner.

\section*{Experiments}

Uniaxial compression tests are performed on 3D-printed, lattice-based metamaterials - consisting of rigid nodes connected by flexible bonds (see Fig.\,\ref{fig:1}\textit{A} and \textit{Materials and Methods}) - using an universal materials tester (Instron 5600), straining the lattice with a constant crosshead speed of $v = \SI{0.5}{\mm \per \s}$, unless mentioned otherwise.
The reaction force $F$ as a function of the applied compression is measured while the structure is observed by a side view camera (Sony alpha 1).
All compression tests are repeated four times, rotating the cubic lattice by 90 degrees around the compression axis after the first and third iteration to capture each face’s dynamical response twice. 

\textbf{Dynamic response.} The dynamical response  to uniaxial compression of two slightly different lattice structures is shown as a sequence of images in Fig.\,\ref{fig:1}\textit{B}.
Making the bonds thicker while keeping the node size the same causes the overall response to transition from local buckling (upper row) to global buckling (lower row) once a critical strain, $\epsilon_{\mathrm{crit}}$, is reached.
During local buckling, the lattice collapses into itself due to counter-rotation of neighboring nodes, which results in an hourglass-like pattern when looked at from the 'front' (here the xz-plane).
In contrast, during global buckling the middle row shears out to one side and all nodes in the same row co-rotate.
With two rotation axes perpendicular to the compression direction, there are four possible rotation directions that can be triggered by the buckling instability, two of which are visible in the images as clockwise and counter-clockwise rotation in the left panels in Fig.\,\ref{fig:1}\textit{B}, which we define as 'front view'.
The rightmost panels in Fig.\,\ref{fig:1}\textit{B} show the corresponding  'side view,' highlighting the difference in self-folding of the five layers as well as the absence of node rotation in the yz-plane. 
\textit{A priori} it is unknown which face of the lattice cube will become 'front' and 'side' during uniaxial compression; however, as long as the time between compression cycles is held short (1 minute in our case) the cube 'remembers' the state set by the first compression.

To highlight the counter-rotating and co-rotating nature of the nodes during both types of buckling,  in Fig.\,\ref{fig:1}\textit{C} the node rotation angle $\theta$ is shown for the center nine nodes of the 'front view'.
Different symbols represent the nodes indicated in the images to the right of each plot.
For both cases we observe that $\theta$ starts to bifurcate above a critical strain $\epsilon_\text{crit}$ and - apart from the middle row during global buckling (green triangles) - $\theta$ increases with $\epsilon > \epsilon_\text{crit}$.  

Since the angles bifurcate during buckling, different regions of the lattice might arbitrarily choose a different rotation pattern, which can result in domain walls.
Figure\,\ref{fig:1}\textit{B} shows a deformation pattern that is uniform throughout the lattice when there is sufficiently strong coupling between nodes by relatively thick or stiff bonds. 
This gives way to non-uniform deformation patterns when the coupling is weak, with domain walls between regions corresponding to differently oriented final states.
Figure\,\ref{fig:1}\textit{D} shows an example for local buckling.
When looked at from the front we see a change in orientation of the pattern at different depths. 
The side view indicates that the left-most layer has a different rotation pattern compared to the other three, implying that the bonds extending across the domain wall (red region) experience torsion.  
Given a large enough system, local buckling can generate domain walls also in lattices with a stronger coupling (see Suppl.\,Info.\,Sec.\,I).

\begin{figure}[t!]
  \centerline{\includegraphics[width=\columnwidth]{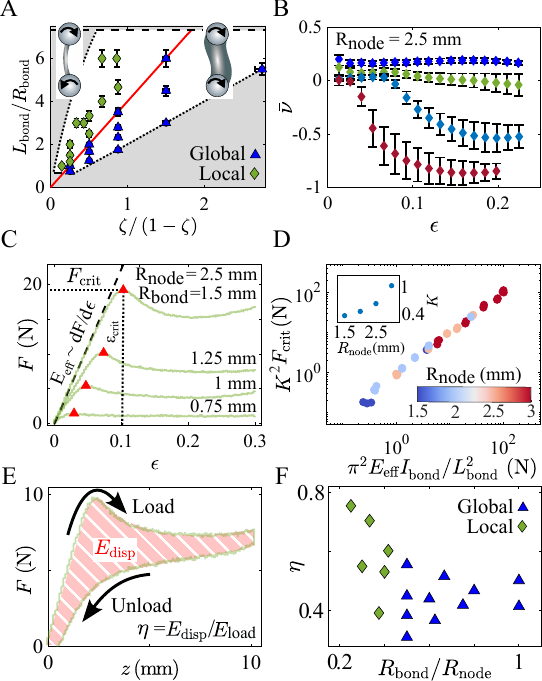}}
\caption{\textbf{Buckling phase space and mechanical response.}
(\textit{A}) Parameter space indicating the regions of global (blue triangles) and local (green diamonds) buckling. The gray area is beyond the available design space. The red line corresponds to Eq.\,\ref{Eq:criterion} and its slope is the ratio of the effective Young's and shear moduli of the lattice material $E_{\mathrm{bond}}/G_{\mathrm{bond}} = 4$. 
(\textit{B}) Poisson's ratio averaged over the center nine nodes $\bar{\nu}$ as a function of applied strain $\epsilon$ for the 'front' view.  Dark-blue to red, $R_{\mathrm{bond}} = [\SI{2}{\mm}, \SI{1.25}{\mm}, \SI{1}{\mm}, \SI{0.75}{\mm}]$. 
(\textit{C}) Reaction force $F$ during compression as a function of strain $\epsilon$ for four different lattice designs with $R_{\mathrm{node}} = \SI{2.5}{\mm}$. 
The first local maximum indicates $F_{\mathrm{crit}}$ and $\epsilon_{\mathrm{crit}}$ (red triangles).
The initial slope (dashed line) gives the effective Young's modulus $E_{\mathrm{eff}} \sim \mathrm{d}F/\mathrm{d}\epsilon$ of the lattice.
(\textit{D}) Normalized critical buckling force of 15 different designs as a function of material properties on a double-logarithmic scale.  
Each point represents one experiment and its color reflects the node radius.
The inset shows the measured relation between $K$ as a function of $R_{\mathrm{node}}$ for these lattices.
(\textit{E}) Typical force versus displacement curve to determine the dissipation energy $E_{\mathrm{disp}}$ (red shaded area) and the dissipation ratio $\eta = E_{\mathrm{disp}} /E_{\mathrm{load}}$. 
(\textit{F}) Dissipation ratio $\eta$ for different designs with $N = 5$ as a function of $R_{\mathrm{bond}}/R_{\mathrm{node}}$. Symbols indicate global (triangles) or local (diamonds) buckling. Standard deviations over three repetitions are within the size of the symbols. 
}
\label{fig:2}
\end{figure}

\textbf{Global versus local buckling.} To map out the conditions under which the response to compression involves global or local buckling, we vary the relative sizes of nodes and bonds as well as the overall lattice size.
From Fig.\,\ref{fig:1}\textit{E} we find that designs with larger $L_{\text{bond}}$ or larger $R_{\text{bond}}$ exhibit uniform global buckling (blue triangles), characterized by shear between successive lattice planes along $z$ and uniform rotation of nodes within each plane.
Once $L_{\text{bond}}$ or $R_{\text{bond}}$ fall below certain threshold values, the deformation pattern switches to local buckling (green diamonds), characterized by alternating bending of the bonds and counter-rotating neighboring nodes within and across layers.  
For sufficiently thin bonds we furthermore observe domain wall formation (red squares), whereby the bonds connecting adjacent vertical planes experience twist.
As we will show below, in the continuum limit the transition from global buckling (facilitated by shear between lattice planes)  to local buckling (facilitated by bond bending) occurs when $4 k_b/(k_s L_{\mathrm{bond}}^2) = 1$. Here the bond bending stiffness $k_b = E_{\mathrm{bond}} I_{\mathrm{bond}}/L_{\mathrm{bond}}$ and the shear stiffness $k_s = \kappa G_{\mathrm{bond}} A_{\mathrm{bond}}/L_{\mathrm{bond}}$ depend on
the effective Young's and shear moduli $E_{\mathrm{bond}}$ and $G_{\mathrm{bond}}$ of the lattice material.  
Translating this result to a criterion for a lattice fabricated from rods as bonds and spheres as nodes, we use the bonds’ area moment of inertia $I_{\mathrm{bond}} = \pi R_{\mathrm{bond}}^4 / 4$ and their cross-sectional area  $A_{\mathrm{bond}} = \pi R_{\mathrm{bond}}^2$. 
To model the effective stiffness to layer shearing, we use a shear correction factor $\kappa$ \cite{timoshenko2012theory} and set $\kappa = \frac{1 - \zeta}{\zeta} \, R_{\mathrm{bond}}/L_{\mathrm{bond}}$  to capture the decrease of effective shearing stiffness when either the node or the bond radii decrease or when the bond length increases. 
Here $\zeta = (N-1) L_{\mathrm{bond}}/L$ is the fraction of side length $L$ occupied by bonds. 
After substituting this for $k_b$ and $k_s$, the transition criterion becomes
\begin{equation}
\frac{L_{\mathrm{bond}}}{R_{\mathrm{bond}}} = \frac{E_{\mathrm{bond}}}{G_{\mathrm{bond}}} \frac{\zeta}{1 - \zeta}.
\label{Eq:criterion}
\end{equation}
According to Eq.\,\ref{Eq:criterion}, the distinct domains that produce global and local buckling can thus be delineated in a plot of bond slenderness, $L_{\mathrm{bond}}/R_{\mathrm{bond}}$ versus the bond-to-node length ratio, $\zeta/(1 - \zeta)$.
This is shown in Fig.\,\ref{fig:2}\textit{A}, which depicts data for the same designs as Fig.\,\ref{fig:1}\textit{E}.
The areas shown in gray fall outside the available design space due to either printer resolution limitations (dashed line) or design restrictions (dotted line).
For the lattice structures investigated here, we find the ratio of the effective Young's and shear moduli of the lattice material to be $E_{\mathrm{bond}}/G_{\mathrm{bond}} = 4$ (red line).

Apart from controlling the type of buckling that will emerge, decreasing the bond radius while keeping the size of the nodes the same  allows us to tune the degree to which a lattice can fold into itself under compression. 
This can be quantified by determining the Poissons's ratio, $\nu = - \epsilon_{x} / \epsilon_{z}$, where $\epsilon_{x}$ and $\epsilon_{z}$ are the strain in the transverse and axial direction, respectively. 
By tracking the positions of the nodes during compression from the 'front view', we can determine the instantaneous Poisson's ratio $\nu_{i,k}(\epsilon)$ of every node and take an average of the nine in the center.
Figure\,\ref{fig:2}\textit{B} shows the evolution of this average, $\bar{\nu}$, as a function of strain for four lattices with constant node size but decreasing bond radius.
By doing so, we transition from global buckling (darkblue and green diamonds) with $\bar{\nu} \gtrapprox 0$ to local buckling (lightblue and red diamonds) with $\bar{\nu} \lessapprox 0$ beyond a critical strain.
Tuning the geometry of the lattice structure thus allows for a broad range of control over the local Poisson's ratio. 

\textbf{Force response.} 
For four different designs with constant node radius but decreasing bond radius, the force-strain relationships are shown in Fig.\,\ref{fig:2}\textit{C}.
Initially linearly increasing with applied strain, $F$ reaches a maximum before either weakening (negative slope) for thicker bonds or plateauing (zero slope) for thinner ones.
The slope of the initial increase relates to the effective Young's modulus of the cubic lattice as $E_{\mathrm{eff}} \sim \mathrm{d}F / \mathrm{d} \epsilon$. 
Per definition $E_{\mathrm{eff}} = \mathrm{d} \sigma / \mathrm{d} \epsilon$, with $\sigma$ the local stress. 
Considering that only the vertical bonds are load-bearing during the initial stages of compression, we write $E_{\mathrm{eff}} = \left(\rm{d}F/\rm{d}\epsilon\right) / \left(N^2A_{\rm{bond}} \right)$.

The reaction force and strain at the first local maximum, $F_{\mathrm{crit}}$ and $\epsilon_{\mathrm{crit}}$, indicate the critical force and strain at which either global or local buckling occurs.
Using the above expression for $E_{\mathrm{eff}}$, the two can be related by $F_{\mathrm{crit}} \approx N^2 E_{\mathrm{eff}} A_{\mathrm{bond}} \epsilon_{\mathrm{crit}}$.
Based purely on this force-strain relationship we cannot tell what type of buckling is triggered. 
Approaching the buckling of the simple cubic lattice as a classical buckling problem of all the load-bearing, vertical bonds, we argue that Euler's critical load criteria should hold \cite{euler1952methodus,timoshenko2012theory}, which balances bending and compression energies.
Hence,
\begin{equation}
F_{\rm{crit}} = \frac{\pi^2E_{\mathrm{eff}}I_{\rm{bond}}}{(KL_{\rm{bond}})^2},
\label{eq:EulerLoad}
\end{equation} 
with $E_{\mathrm{eff}}$ as defined above, $I_{\mathrm{bond}} = \pi R_{\mathrm{bond}}^4/4$ the area moment of inertia of the bond, and $K$ a dimensionless pre-factor of order unity that effectively changes the lengths of the bond and relates to the applied boundary conditions \cite{timoshenko2012theory}.
As we leave the system unconstrained, apart from compression in one direction, the top and bottom boundary nodes are essentially free to rotate and translate in the transverse directions. 
Since it is unknown how the overall geometry of the lattice will affect the local boundary conditions, and hence the value of $K$, we measure $F_{\mathrm{crit}}$ experimentally and relate this to the known material properties - in line with Eq.\,\ref{eq:EulerLoad} - to empirically determine this pre-factor for different designs.
A collapse of the data is achieved and shown on a double-logarithmic scale in Fig.\,\ref{fig:2}\textit{D} after normalization with $K$ (see inset Fig.\,\ref{fig:2}\textit{D}).
Knowing $K(R_{\mathrm{node}})$ for a given set of designs then allows us to predict the critical force (and hence critical strain) at which buckling will occur, given a specific set of lattice material properties. 

\textbf{Energy dissipation.} 
During a loading and unloading cycle the lattice structures will dissipate a certain amount of energy.
To investigate the design-dependent strain energy dissipation we perform compression cycles for a subset of the designs with $N=5$ (see \textit{Materials and Methods}) and determine the dissipation ratio $\eta = E_{\mathrm{disp}} /E_{\mathrm{load}} = 1 - \vert E_{\mathrm{unload}} \vert /E_{\mathrm{load}}$, based on the force-displacement curves, see Fig.\,\ref{fig:2}\textit{E}.
We then quantify for each design the strain energy dissipation in Fig.\,\ref{fig:2}\textit{F}.
We find that designs exhibiting local buckling (diamonds) - and specifically those with thin bonds and large nodes - are more efficient in terms of strain energy dissipation \cite{yin2024quasi} as compared to those structures that exhibit global buckling (triangles).

Our lattice-based metamaterials thus facilitate a platform where node rotation and bond bending during local buckling favour energy dissipation.
Being able to nucleate and steer local buckling within such lattice structures in a controllable and predictable manner
is the focus of the remainder of this paper.   

\section*{Continuum model}

In order to extend and generalize our experimental observations to larger systems, we aim to construct a mathematical framework allowing us to approach the continuum limit \cite{deng2017elastic,deng2018metamaterials,mo2019cnoidal,deng2020characterization,deng2021dynamics,veenstra2024non}.
Specifically, we wish to describe the temporal evolution of the node rotations within the 3D lattice during compression.

\begin{figure}[t!]
  \centerline{\includegraphics[width=\columnwidth]{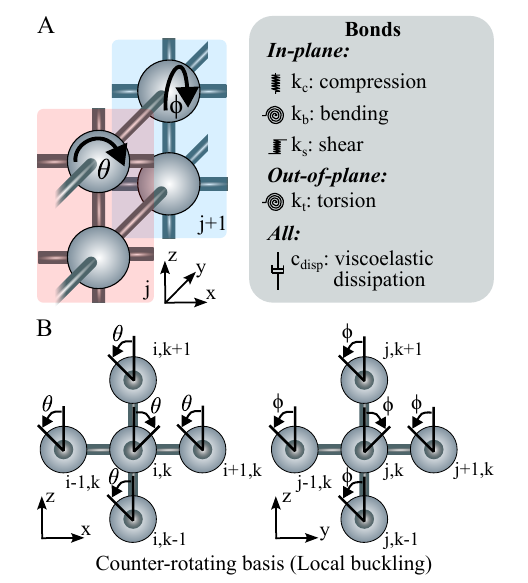}}
\caption{\textbf{Definitions for 3D-continuum model.}
(\textit{A}) 3D-schematic indicating different xz-planes (red and blue regions) and the corresponding $\theta$ and $\phi$ node rotations in the xz- and yz-plane, respectively. 
The box indicates bonds of the cubic lattice that are modelled as axial (compression and shear) or torsional springs (bending and torsion). 
All bonds experience viscoelastic dissipation.
(\textit{B}) Schematic of a sliced portion of the cubic lattice showing the definitions for $\theta$ and $\phi$ in their corresponding planes, as well as the indexing of the nodes.
We assume a counter-rotating coordinate transformation within the plane, resembling local buckling.
}
\label{fig:3}
\end{figure}

To this end, we first consider a discrete set of nodes that are connected with flexible bonds.
Each node has a rotation angles $\theta_{i,j,k}$ and $\phi_{i,j,k}$ associated within xz- and yz-planes, respectively, and we apply a coordinate transformation in line with the experimental observations on counter-rotating neighbours during local buckling, see Fig.\,\ref{fig:3}.
This entails that if $\theta_{i,j,k}$/$\phi_{i,j,k}$  is defined positive for clockwise rotation,  its in-plane neighbours are defined positive for counter-clockwise rotations. 
We then consider that the connecting bonds act as harmonic springs.
For in-plane bonds we assume axial springs for compression and shear, and torsional springs for bending.
For out-of-plane bonds we assume an additional torsional spring for torsion, connecting nodes of different planes. 
For all bonds we lastly assume viscoelastic dissipation, see Fig.\,\ref{fig:3}\textit{A}. 
The current model only accounts for node rotations during compression and no translations.
The latter are related to the applied axial strain ($\epsilon_z$), as well as $\epsilon_{x}$ and $\epsilon_{y}$, that relate to the local Poisson's ratio, which is a lattice structure property (see Fig.\,\ref{fig:2}\textit{B}).
Here, we assume that the 3D strains are known and put ourselves in the reference frame of the translating nodes, hence only considering rotation as a degree of freedom.
In addition, we impose a constraint on the maximum rotation angle, in line with the experimental observations of nodal contact at sufficiently large strain.

We then have corresponding energies in 1D for bending ($E_b = k_b/2 \sum_k \left[\theta_k + \theta_{k+1}\right]^2 +  \left[\phi_k + \phi_{k+1}\right]^2$), shearing ($E_s = \left(L_{\mathrm{bond}}/2\right)^2 k_s /2 \sum_k \left[\theta_k - \theta_{k+1}\right]^2 +  \left[\phi_k - \phi_{k+1}\right]^2$), compression ($E_c = -F_z L_{\mathrm{bond}} \sum_k \left[1 - \cos \theta_{k} \cos \phi_{k} \right]$) and a Rayleigh dissipation function \cite{landau1960mechanics, ginsberg1998advanced} ($\mathcal{F}_{\theta}(\dot{\theta}) = c_{\mathrm{disp}}/2 \sum_k \dot{\theta}_k^2$),
where $k$ goes from $1$ to $N$ and $\dot{\theta} = \partial_t \theta(\vec{x},t)$ with $\vec{x} = [x,y,z]$.
The 3D equivalents of all expressions are listed in the Suppl.\,Info.\,Sec.\,II.
$F_z$ is the applied force defined positive in the negative z-direction, and $c_{\mathrm{disp}}(\dot{\epsilon})$ is the strain-rate dependent dissipation factor (see \textit{Materials and Methods}).
The torsional energy connecting 2D planes is $E_t = k_t/2 \sum_{i,j,k} \left[\theta_{i,j,k} - \theta_{i,j+1,k}\right]^2 +  \left[\phi_{i,j,k} - \phi_{i+1,j,k}\right]^2$, where $k_t = G_{\mathrm{bond}}J_{\mathrm{bond}}/L_{\mathrm{bond}}$, with $J_{\mathrm{bond}} = \pi R_{\mathrm{bond}}^4/2$ the bond's polar moment of intertia. 
Inertial effects are neglected since we assume to be in the overdamped limit.
Its potential effect is discussed in detail in the Suppl.\,Info.\,Sec.\,VII, including a dynamical stability analysis and topological defects solutions.

Next, we evaluate a simplified Lagrangian formulation \cite{landau1960mechanics} $\partial_{\theta_{i,j,k}} (E_b + E_s + E_c + E_t)  + \partial_{\dot{\theta}_{i,j,k}} \mathcal{F}_{\theta} = 0$ (equivalently for $\phi_{i,j,k}$) and assume expansions for $\theta_{k \pm 1} \approx \theta \pm L_{\mathrm{bond}} \, \partial_{z} \theta + L_{\mathrm{bond}}^2/2 \, \partial^2_z \theta$ (equivalently for $\theta_{i+1}$ and $\theta_{j+1}$) to approach the continuum limit.
Additionally, we approximate the angles, using $\sin \theta \approx \theta - \theta^3/6$ and $\cos \theta \approx \theta - \theta^2/2$.  
After evaluation and rearrangement of terms (see Suppl.\,Info.\,Sec.\,II for a detailed derivation) we obtain the following set of reaction-diffusion equations:
\begin{equation}
\begin{aligned}
\partial_t \theta(\vec{x},t) = & D_{xz} \left(\partial_x^2 + \partial_z^2 \right) \theta(\vec{x},t) + D_{y}  \partial_y^2 \theta(\vec{x},t) + f(\theta,\phi), \\
\partial_t \phi(\vec{x},t) =  & D_{yz} \left(\partial_y^2 + \partial_z^2 \right) \phi(\vec{x},t) + D_{x}  \partial_x^2 \phi(\vec{x},t) + g(\theta,\phi),
\end{aligned}
\label{eq:3DReactionDiffusion}
\end{equation}
with $D_{xz} = D_{yz} = \left(k_s L_{\mathrm{bond}}^2/4 - k_b \right)L_{\mathrm{bond}}^2 / c_{\mathrm{disp}}$ and $D_y = D_x = k_t L_{\mathrm{bond}}^2 / c_{\mathrm{disp}}$ the diffusivity constants in the xz- and yz-planes, and y- and x-directions, respectively. 
The forcing terms read:
\begin{equation}
\begin{aligned}
f(\theta,\phi) = & \mathcal{C}_{\theta} \left[1  - \frac{8 k_b}{F L_{\mathrm{bond}}}  - \frac{\theta(\vec{x},t)^2}{6} - \frac{\phi(\vec{x},t)^2}{2} \right] \theta(\vec{x},t),  \\
g(\theta,\phi) = & \mathcal{C}_{\phi}  \left[1  - \frac{8 k_b}{F L_{\mathrm{bond}}}  - \frac{\phi(\vec{x},t)^2}{6} - \frac{\theta(\vec{x},t)^2}{2} \right] \phi(\vec{x},t),  
\end{aligned}
\label{eq:3DReactionDiffusion_SourceTerms}
\end{equation}
where $\mathcal{C}_{\theta} = \mathcal{C}_{\phi} = F_z L_{\mathrm{bond}} / c_{\mathrm{disp}}$. 
Repeating the derivation in the co-rotating reference frame (resembling global buckling; see Suppl.\,Info.\,Sec.\,II) yields a similar set of equations but with $D_{xz}^{\star} = D_{yz}^{\star} = \left(k_b - k_s L_{\mathrm{bond}}^2/4 \right)L_{\mathrm{bond}}^2 / c_{\mathrm{disp}}$.  The crossover between local and global buckling occurs when the diffusivities associated with these two modes become equal, i.e., when  $\left(k_b - k_s L_{\mathrm{bond}}^2/4 \right) = \left(k_s L_{\mathrm{bond}}^2/4 -k_b \right)$, which produces the transition criterion at $4 k_b/(k_s L_{\mathrm{bond}}^2) = 1$, that we used earlier in the derivation of Eq.\,\ref{Eq:criterion}.

\textbf{Static buckling.} In the static limit we can analytically determine the buckling behavior considering $f(\theta,\phi=0) = 0$, resulting in non-trivial solutions of the form $\theta = \pm \left( 6 \left[1 - 8 k_b / (F_z L_{\mathrm{bond}}) \right] \right)^{1/2}$. 
The solution for $\theta$ bifurcates at a critical force - similar to that in Eq.\,\ref{eq:EulerLoad} of the whole lattice structure - where now $F_{\mathrm{crit}} = 8 k_b/L_{\mathrm{bond}}$. 
As earlier, $F = N^2 E_{\mathrm{eff}} A_{\mathrm{bond}} \epsilon$ in the initial linear regime (see \textit{Materials and Methods} for a more general alternative), hence $\epsilon_{\mathrm{crit}} = 8 k_b / (N^2 E_{\mathrm{eff}} A_{\mathrm{bond}} L_{\mathrm{bond}}) = 8 k_b / ( N^2 k_c L_{\mathrm{bond}}^2)$.
This leads to $\theta  =\pm \left( 6 \left[1 - \epsilon_{\mathrm{crit}} / \epsilon \right] \right)^{1/2}$ with bifurcation occurring at $\epsilon = \epsilon_{\mathrm{crit}}$ and with $\theta$ increasing with applied strain - for both local and global buckling - thus in line with the experimental observations in Fig.\,\ref{fig:1}\textit{C}. Using this, we can rewrite Eq.\,\ref{eq:3DReactionDiffusion_SourceTerms} in terms of strain as
\begin{equation}
\begin{aligned}
f(\theta,\phi) = & \mathcal{R}_{\theta} \epsilon \left[1  - \epsilon_{\mathrm{crit}} / \epsilon  - \frac{\theta(\vec{x},t)^2}{6} - \frac{\phi(\vec{x},t)^2}{2} \right] \theta(\vec{x},t),  \\
g(\theta,\phi) = & \mathcal{R}_{\phi} \epsilon  \left[1  - \epsilon_{\mathrm{crit}} / \epsilon  - \frac{\phi(\vec{x},t)^2}{6} - \frac{\theta(\vec{x},t)^2}{2} \right] \phi(\vec{x},t),  
\end{aligned}
\label{eq:3DReactionDiffusion_SourceTerms_strain}
\end{equation}
where $\mathcal{R}_{\theta} = \mathcal{R}_{\phi} = N^2 k_c L_{\mathrm{bond}}^2 / c_{\mathrm{disp}}$. 
Interestingly, the force terms change sign as we increase the applied strain and transitions from being negative at small strains (sinks) to becoming positive (sources) when $\epsilon \gtrsim \epsilon_{\mathrm{crit}}$.
As a consequence, initial perturbations in the node rotations decay prior to buckling, but grow during post-buckling. 

\textbf{Buckling plane selection.} As a next step we analyze the stability of the system when both $\theta$ and $\phi$ are non-zero to determine why node rotations and therefore buckling are only observed in one of the planes (see Fig.\,\ref{fig:1}\textit{B}).
We assume a potential energy $V(\theta,\phi)$ for which $\partial V(\theta,\phi)/\partial \theta = - c_{\mathrm{disp}} f(\theta,\phi) = 0$ and $\partial V(\theta,\phi)/\partial \phi = - c_{\mathrm{disp}} g(\theta,\phi) = 0$ need to be satisfied. 
If both variables are non-zero they must be equal, leading - after subtraction of the terms in brackets of Eq.\,\ref{eq:3DReactionDiffusion_SourceTerms_strain} - to $\theta^2 = \phi^2 = \xi^2 = 3\left(1 - \epsilon_{\mathrm{crit}} / \epsilon \right)/2$.
Integration yields a potential energy as
\begin{equation}
V =- c_{\mathrm{disp}} \mathcal{R} \epsilon \left[ \left( 1 - \frac{\epsilon_{\mathrm{crit}}} { \epsilon} \right)\frac{ \theta^2 + \phi^2 }{2} - \frac{\theta^4 + \phi^4}{24} - \frac{\theta^2 \phi^2}{4} \right],
\end{equation}
with $\mathcal{R}=\mathcal{R}_{\theta}=\mathcal{R}_{\phi}$, resembling the Landau free-energy of a second-order phase transition \cite{chaikin1995principles}.
Determining its Hessian $H$ and substituting $\theta^2 = \phi^2 = \xi^2$ yields that $\det(H) \propto -2 \left[ \left(1 - \epsilon_{\mathrm{crit}} / \epsilon\right) \right]^2 < 0$ (see Suppl.\,Info.\,Sec.\,II), which is unstable. 
However, when $\theta \neq 0$ and $\phi = 0$ or vice versa, the Hessian is diagonal with positive diagonal elements, representing a stable minimum. 
This is therefore in line with our experimental observations that only one face of the cubic lattice shows node rotation and pattern formation (front view), whereas rotations are absent on the perpendicular faces (side view), see Fig.\,\ref{fig:1}\textit{B}. 

\begin{figure}[t!]
  \centerline{\includegraphics[width=\columnwidth]{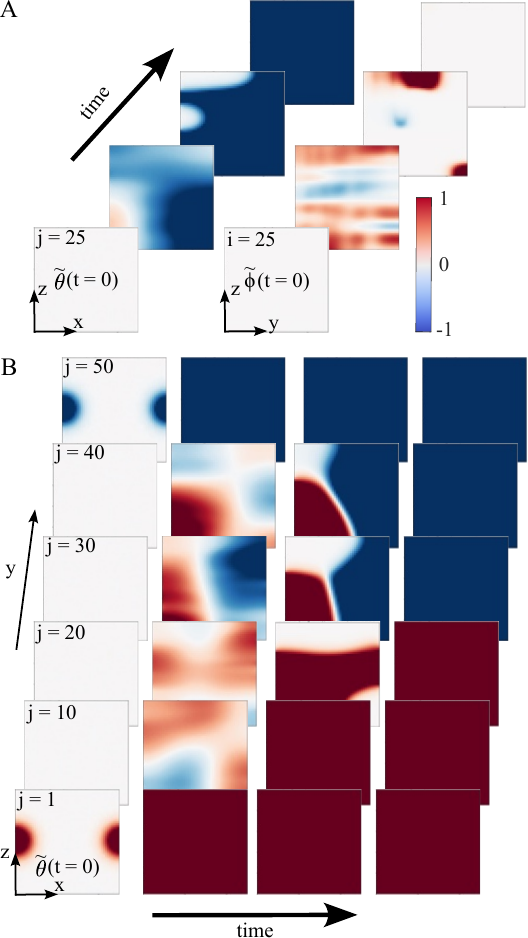}}
\caption{\textbf{Buckling plane selection and propagation in 3D.}
(\textit{A}) Cross-sections of the 3D domain showing the temporal evolution of the normalized node rotations $\tilde{\theta} = \theta/\theta_{\mathrm{max}}$ and $\tilde{\phi} = \phi/\phi_{\mathrm{max}}$ in the counter-rotating basis, given random initial angular perturbations.  
Eventually, a buckling plane is selected with $\tilde{\theta} \neq 0$ and $ \tilde{\phi} = 0$, or vice versa.
(\textit{B}) Local buckling propagation through a 3D lattice structure given nucleation sides at the less constrained edges and random initial angular perturbations. 
}
\label{fig:4}
\end{figure}

\textbf{Numerical evaluation.} 
The evolution of the node rotations can now be determined by numerically evaluating Eqs.\,\ref{eq:3DReactionDiffusion}\,\&\,\ref{eq:3DReactionDiffusion_SourceTerms_strain} (see \textit{Materials and Methods}).
Starting with a randomly perturbed initial state, Fig.\,\ref{fig:4}\textit{A} shows the temporal evolution of normalized angles $\tilde{\theta} = \theta/\theta_{\mathrm{max}}$ and $\tilde{\phi} = \phi/\phi_{\mathrm{max}}$ within two fixed planes in the counter-rotating basis.
Here, $\theta_{\max} = \phi_{\max} = \SI{40}{\degree}$ is the maximum rotation angle of the nodes observed experimentally.
As time progresses, the above-mentioned buckling plane selection ensures that $\tilde{\theta} \neq 0$ and $\tilde{\phi} = 0$ for this case. 
Next, we can impose an additional initial condition, consistent with our experimental observations, that buckling nucleation sides occur at the least constrained edges of the cubic lattice.
We then follow the propagation of the buckling instability through the different layers of the 3D system, leading to the series of snapshots shown in Fig.\,\ref{fig:4}\textit{B}. 
Over time a static xz-plane emerges in which the orientation of the local buckling completely changes (from red to blue), indicating the existence of a domain wall (last row Fig.\,\ref{fig:4}\textit{B}).
By varying the strength of the initial perturbation and/or the coupling strength between planes - set by $D_x$ and $D_y$ - the position of this domain wall can shift or can become absent, potentially leading to uniform local buckling. 
Similar behavior is observed in the experiments, see Fig.\,\ref{fig:1}\textit{B}\,\&\,\textit{E}. 

\begin{figure}[p]
  \centerline{\includegraphics[width=\columnwidth]{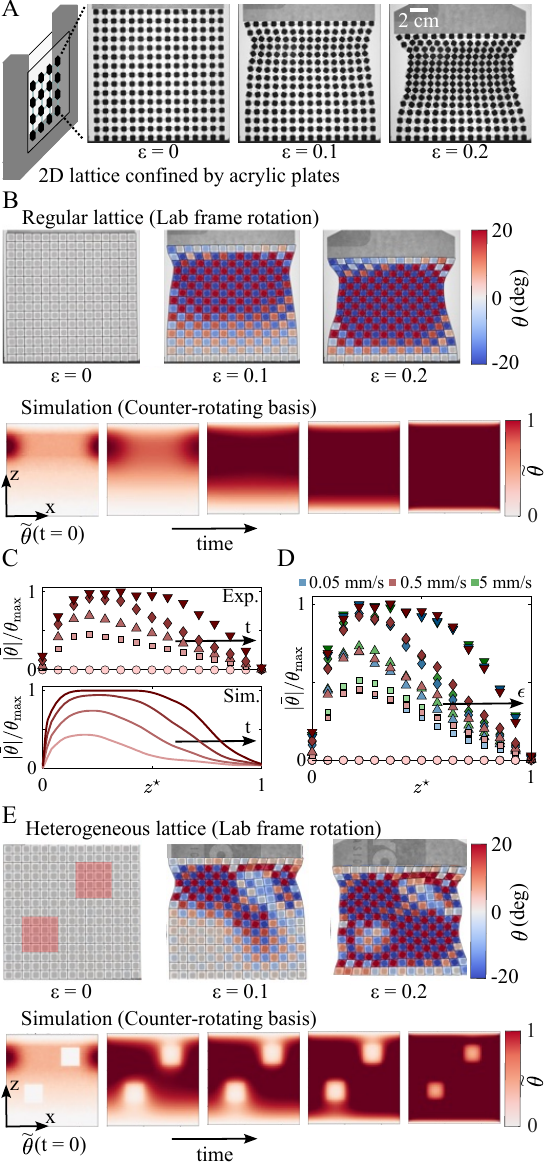}}
  \caption{\textbf{Diffusive buckling fronts in 2D lattices.}
 (\textit{A}) Sketch of setup and sequence of images during compression, see Suppl.\,Movie\,3\,\&\,4. 
(\textit{B}) Top: Overlayed node lab frame rotation during compression of the regular lattice.
Bottom: 2D simulations with similar initial conditions showing the propagating buckling front in the counter-rotating basis.  
(\textit{C}) Normalized averages of the rotation angles as a function of rescaled vertical coordinate $z^{\star}$ at different instances in time, for crosshead speed $\SI{0.5}{\mm \per \s}$.
(\textit{D}) Same as (\textit{C}) but for three different crosshead speeds, collapsed by normalizing time by strain rate. Different traces are for different strain. 
(\textit{E}) Same as (\textit{B}) but with thicker bonds in the red regions. }
\label{fig:5}
\end{figure}

\textbf{Diffusive buckling front propagation} 
The above analysis indicates that diffusive fronts will propagate within 2D planes of the 3D lattice structure.  
Especially if the coupling between the different layers is strong (weak) the dynamics within the plane parallel to the compression axis become identical (isolated).
To track this propagation we therefore can use a representative 2D lattice system and, for comparison with the continuum model, scale up the number of nodes to increase resolution. 
This is shown in Fig.\,\ref{fig:5}\textit{A}-\textit{B} for a square lattice with 15x15 nodes (see \textit{Materials and Methods}).
We confine the 2D lattice between two acrylic plates spaced $\SI{6}{\mm}$ apart to ensure $\phi = 0$ (see sketch in Fig.\,\ref{fig:5}\textit{A})  - hence biasing the buckling plance selection - and compress the structure, given a target strain and a fixed crosshead speed of $v = \SI{0.5}{\mm \per s}$.
A corresponding sequence of images is shown in Fig.\,\ref{fig:5}\textit{A}.
Similar to the three-dimensional cubic lattices, we observe counter-rotation of the nodes, where - due to the confinement - buckling is consistently initiated at the upper edges of the lattice (see \textit{Materials and Methods}).
For better indication we have overlayed the experimental footage with the determined local node rotation in Fig.\,\ref{fig:5}\textit{B}, highlighting counter-rotation, as well as the formation of a buckling front propagating first inwards and then downwards (see Suppl.\,Movie\,3).
The bottom series of snapshots shows the results after evaluating Eq.\,\ref{eq:3DReactionDiffusion}\,\&\,\ref{eq:3DReactionDiffusion_SourceTerms_strain}, assuming fixed top and bottom boundaries, and with $\phi = D_y = 0$, as well as initial conditions similar to those in the experiments (see squares in Fig.\,\ref{fig:5}\textit{C}), showing qualitative agreement. 

For better comparison, we take the average of the absolute node rotations at each vertical position, and plot its normalised value as a function of dimensionless and rescaled vertical position $z^{\star} = 1 - z/ \left(L \left[1 - \epsilon \right] \right)$, going from top to bottom and taking compression into account, see Fig.\,\ref{fig:5}\textit{C}.
The plots show the evolution of the experimentally observed (top) and numerically simulated (bottom) buckling front with different symbols/colours indicating different instances in time (or strain with $\epsilon = [0.05, 0.1, 0.15, 0.2]$), again showing satisfactory agreement.
A quantitive analysis, comparing front location, width and velocity, is in Suppl.\,Info.\,Sec.\,V. 
Given the viscoelastic nature of the polyjet-printed resin (Agilus 30) \cite{abayazid2020material} its material properties are strain-rate dependent. 
However, over a large range of compression speeds ($\SI{0.05}{\mm \per \s} \leq v \leq \SI{50}{\mm \per \s}$) the dynamics in our experiments remain unaltered when normalizing time by strain rate (see Fig.\,\ref{fig:5}\textit{D} and \textit{Materials and Methods}).

\textbf{Front manipulation.} An important outcome of our model is the direct link between lattice design parameters and buckling front propagation. This provides a means for manipulating and steering the fronts. Here we discuss two examples where this is achieved through local changes in the lattice structure, although changing local material parameters, such as the bond's Young's modulus, provides an alternate approach.
In Fig.\,\ref{fig:5}\textit{E}, bonds within the red regions have been made thicker ($R_{\mathrm{bond}} = \SI{1.5}{\mm}$), causing local variations in i) the diffusivity constant ($D_{xz}$), ii) the strength of the source term ($R_{\theta}$), and iii) the onset of the latter becoming positive ($\epsilon_{\mathrm{crit}}$). 
This leads to front propagation around these regions - as predicted by the dynamic buckling model by assuming different coefficients locally - followed by their delayed buckling onset.
Steering is also achieved by generating structural defects, such as randomly misplaced bonds, within suitably identified regions, which retards node rotation and causes the formation of (temporary) internal boundaries (see Suppl.\,Movie\,3\,\&\,4).
The same methodology might now be applied to 3D-networks allowing for an expanded design space to enhance instability propagation, shielding of certain regions, or potentially (local) force attenuation - all of which are future research directions still to be explored.
The latter might be of particular interest as we find that the amount of energy dissipated during one loading cycle is strongly dependent on lattice structure (see \textit{Materials and Methods} and Suppl.\,Info.\,Sec.\,III).

\section*{Conclusions}

Investigating the dynamic buckling behavior under uniaxial compression of 3D-printed, regular lattice-based metamaterials comprised of flexible bonds connecting rigid nodes,
we mapped the design space that controls different buckling modes, delineating the transition from global to local buckling.  In these dissipative, overdamped structures we find that local buckling, triggered by exceeding a threshold strain, propagates as diffusive front. 
Increasing the bond slenderness changes the  response to applied strain from uniform local buckling patterns to non-uniformly buckled configurations, where weak coupling accommodates domain walls between regions corresponding to differently oriented configurations.
Bond slenderness also controls the strain-dependent Poisson's ratio, making it possible to move from normal elastic to deeply auxetic behavior.

A 3D-continuum model for the evolution of the node rotations during buckling in overdamped, lattice-based metamaterials leads to a  set of coupled reaction-diffusion equations.
Both forcing terms change sign with increasing applied strain, transitioning from being sinks to source terms that enhance node rotation.
The model, validated by our experiments, captures the onset and the propagation of buckling instabilities through coefficients that depend on material properties as well as local geometry.  
We show that this opens up opportunities for deliberate manipulation and control of cascading buckling events, for example by designing heterogeneities into the lattice. 
Since local bond bending and node rotation are essential for the formation of diffusive buckling front that allow for controlled manipulation, lattice designs should not be too restrictive for node rotations.
We expect that this approach could be reversed and used to extract the local material properties in a given metamaterial, in particular when combined with physics-based machine learning \cite{hernandez2025data}.  While our experiments and modeling apply to the regime where material properties can be taken as constant, the viscoelastic behavior of polyjet-printed resin should, in principle, also make it possible to enter regimes where the dynamical response of the metamaterial becomes trainable \cite{jaeger2024training}.  In that case, repeated forcing via an appropriately chosen training protocol could condition the structures to respond in a targeted manner \cite{gowen2025training,gowen2025training2}. 

\section*{Materials and Methods}

\subsection*{Experiments}
The lattice-based metamaterials are fabricated using a high-resolution polyjet 3D-printer capable of multi-material deposition (Stratasys J850).
The structures are lattices with rigid nodes (VeroWhite Polyjet Resin) connected by flexible bonds (Agilus 30 Polyjet Resin, Shore-A40), see Fig.\,\ref{fig:1}\textit{A}.
Although the mechanical response is dependent on the loading-rate, the most important feature for our study, namely the dynamics of the node rotations, remain unchanged upon proper normalisation for compressions up to $v_{\mathrm{crosshead}} = \SI{50}{\mm \per \second}$ (see Suppl.\,Info.\,Sec.\,IV).
For the 3D simple cubic lattice, we restrict ourselves to $N = \{4, 5\}$ number of nodes per side (hence 64 or 125 nodes in total), inner lattice size $\SI{18}{\mm} \leq L \leq \SI{33.75}{\mm}$, node radius $\SI{1}{\mm} \leq R_{\mathrm{node}} \leq \SI{3.35}{\mm}$ and bond radius $\SI{0.75}{\mm} \leq R_{\mathrm{bond}} \leq \SI{2}{\mm}$  (see Fig.\,\ref{fig:1}\textit{A}). 
The bond length is set by the geometrical relation $L_{\mathrm{bond}} = \left[ L - (2N-2)R_{\mathrm{node}} \right] / \left[ N-1 \right]$ and the printer resolution is $\pm \SI{50}{\micro \meter}$.
Imperfections during printing with respect to bond thickness and stiffness might bias the location of the domain wall, yet the buckling plane selection is found to be robust against local and small imperfections caused during 3D printing.

For the loading and unloading experiments we compress the structure to a target strain of $\epsilon = 0.3$ with crosshead speed $v = \SI{0.5}{\mm \per \s}$, pause for $\SI{5}{\second}$, and unload at the same velocity. 
The mechanical work associated with the force-displacement cycle is calculated by integrating the force over displacement, $E = \int_{\mathcal{\delta}} F(z)\,\mathrm{d}z$
where $\mathcal{\delta}$ denotes the relevant loading or unloading path.
By construction, $E_{\mathrm{load}}>0$ and $E_{\mathrm{unload}}<0$. 
The dissipated energy, corresponding to the area enclosed by the force-displacement loop, is then $E_{\mathrm{disp}} = E_{\mathrm{load}} - \vert E_{\mathrm{unload}}\vert$.
Further details are provided in Suppl.\,Info.\,Sec.\,III.

For the 2D experiments, we consider a 15x15 lattice where the spherical nodes are replaced by squares with side lengths of $\SI{6}{\mm}$ and the bonds have radius $R_{\mathrm{bond}} = \SI{1}{\mm}$.
The length of the entire lattice is $L = \SI{14}{\cm}$.
Compressions are recorded using a camera (Sony alpha 1) and back-lighting. 
Due to confinement by the acrylic plates in the 2D setup (see Fig.\,\ref{fig:5}\textit{A}) we observe that buckling is consistently initiated at the upper edges of the lattice. The reason for this is the friction between the nodes and the acrylic plates, which introduces a slight gradient in the strain field along the z-direction.
Since $\theta(\vec{x},t)  =\pm \left( 6 \left[1 - \epsilon_{\mathrm{crit}} / \epsilon(\vec{x},t) \right] \right)^{1/2}$, the nodes in the upper rows reach the critical strain first and start to buckle.
Additional design configurations are discussed in Suppl.\,Info.\,Sec.\,VI. \\

\subsection*{Transition between force- to strain-driven buckling}
In the derivation of the continuum model we assumed a linear relation between the applied force and the experienced strain, through $F = N^2 E_{\mathrm{eff}} A_{\mathrm{bond}} \epsilon$, as a simplified approach to transition from a force- to a strain-driven system.
More generally, any functional relation where $F = h(\epsilon)$ can be substituted in Eq.\,\ref{eq:3DReactionDiffusion_SourceTerms} to more closely approximate the experiments (see Fig.\,\ref{fig:2}\textit{C}). \\

\subsection*{Numerical evaluation}
We numerically evaluate the dimensionless versions of  Eqs.\,\ref{eq:3DReactionDiffusion}\,\&\,\ref{eq:3DReactionDiffusion_SourceTerms_strain} through a first-order finite difference method with Neumann boundary conditions at the sides and fixed boundaries at top and bottom, in a domain with dimensions $L_x = L_y = L_z = 50$ (see Suppl.\,Info.\,Sec.\,IV).
For modeling our strain-driven experiments we set $\epsilon = \epsilon_0 + \dot{\epsilon} t$ with an initial value for $\epsilon_0$ that can be lower or equal to $\epsilon_{\mathrm{crit}}$, where typically $\epsilon_{\mathrm{crit}} = 0.05$.
The strain rate is chosen to match experiments, with $\dot{\epsilon} = \SI{3e-3}{}$ as typical value.
Through matching of the profiles in Fig.\,\ref{fig:5}\textit{C} we find $D_{xz} / L_{\mathrm{bond}}^2 = \mathcal{O}(1)$ and $\mathcal{R} = \mathcal{O}(1)$ (see Suppl.\,Info.\,Sec.\,IV). \\

\subsection*{Strain-rate dependence} 
The dissipation factor $c_{\mathrm{disp}}(\dot{\epsilon})$ is a function of strain rate $\dot{\epsilon}$.
In particular, $c_{\mathrm{disp}}(\dot{\epsilon}) = \eta_{\mathrm{eff}}(\dot{\epsilon})V_{\mathrm{bond}}$ \cite{landau1960mechanics, landau2012theory}, with $\eta_{\mathrm{eff}}(\dot{\epsilon})$ the effective viscosity of the system.  
For viscoelastic materials experiencing oscillatory strain $\eta_{\mathrm{eff}} = E''(\omega)/\omega$ \cite{landau2012theory}, where $\omega$ is the oscillation frequency and $E''(\omega)$ the frequency dependent loss modulus. 
Hence, it is reasonable to assume that in our case  $\eta_{\mathrm{eff}} \propto E''(\dot{\epsilon})/\dot{\epsilon}$. 
To test whether changes in loss modulus $E''$ are significantly affecting the dynamics when we vary the strain rate, we  performed additional experiments on the 2D lattice at $v = [\SI{0.05}{\mm \per \s},  \SI{5}{\mm \per \s}]$, i.e an order of magnitude smaller and larger than for the data in Fig.\,\ref{fig:5}\textit{C}.
At identical instances of strain, obtained by normalizing time with the corresponding strain rates, we observe a collapse of the experimental data, see Fig.\,\ref{fig:5}\textit{D}, where  curves start to deviate slightly only at high strain rates and at initial times below the critical point. 
Over the range tested in our experiments we can therefore take $E''$ as approximately constant and $c_{\mathrm{disp}}(\dot{\epsilon}) \propto E'' V_{\mathrm{bond}}/\dot{\epsilon}$.
This means that decreasing the strain rate $\dot{\epsilon}$ simply increases the effective viscosity $\eta_{\mathrm{eff}}$ and slows down the dynamics of the system, while our continuum model remains valid.
When the strain rate exceeds the range tested here the loss modulus might change.
In addition, material properties like the bond's Young's and shear moduli, as well as the onset of buckling might vary at higher strain rates, and eventually inertial effects would need be considered.
A discussion is provided in Suppl.\,Info.\,Sec.\,IV\,\&\,VII.\\

\section*{Acknowlegdements}
This work was supported by the Army Research Office under award W911NF-24-2-0184.   The University of Chicago Materials Research Science and Engineering Center, which is supported by the National Science Foundation under award DMR-2011854, provided additional support for sample fabrication through its shared experimental facilities and for F.S. through its Research Experience for Undergraduates summer program. 

\bibliography{References}

\onecolumngrid
\newpage
\section*{Supplementary material}

\section*{List of supplementary movies}

\begin{itemize}
\item Front view of two simple cubic lattices under compression showing local (left) and global (right) buckling, corresponding to Fig.\,1\textit{B}. Timestep indicates passed real time.
\item Front- and side view of a simple cubic lattice under compression showing non-uniform local buckling and the formation of a domain wall (see Fig.\,1\textit{E}). Timestep indicates passed real time.
\item 2D lattice structure under compression, corresponding to Fig.\,5. First panel: Raw experimental data for a regular lattice. Remaining panels: Overlayed nodal rotation for the (second panel) regular lattice, (third panel) irregular lattice, and (last panel) heterogeneous lattice. For the irregular lattice bonds within the two red regions indicated in Fig.\,5\textit{E} have been randomly displaced. Timestep indicates passed real time.
\item 2D simulations corresponding to different cases shown in Fig.\,5.
\end{itemize}

\section*{I - Domain walls in larger lattice structures }

As addressed in the main text, all nodes have two rotation axes that are perpendicular to the compression direction, resulting in four possible rotation directions that can be triggered by the buckling instability.
Since during buckling the angles bifurcate, different layers of the cubic lattice might arbitrarily choose a different orientation as compared to its adjacent layers.
Uniform local buckling (strong coupling; all layers have the same orientation) gives way to non-uniformly buckled configurations (weak coupling), with domain walls between regions corresponding to different final states, see Fig.\,1\textit{E} in the main text.
When looked from the front we see a change in orientation of the pattern-formation at different depths during local buckling. 
The side view clearly indicates that one layer has a different orientation as compared to the other three (green regions), which are separated by a domain wall (red region), that are bonds connecting these layers experiencing torsion.  

\begin{figure}[t!]
  \centerline{\includegraphics[width=\textwidth]{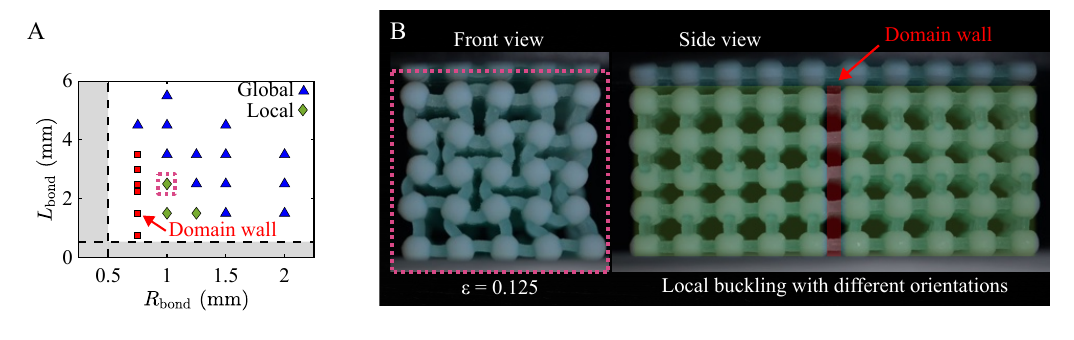}}
\caption{\textbf{Domain walls in extended lattices.}
(\textit{A}) Parameter space in terms of bond length and bond radius indicating the cubic lattices (red squares) that show the formation of a domain wall during local buckling.
(\textit{B}) Front- and side view of a compressed 10x5x5 lattice structure with $R_{\mathrm{node}} = \SI{2.5}{\mm}$, $R_{\mathrm{bond}} = \SI{1}{\mm}$, and $L_z = \SI{30}{\mm}$.  Although strongly coupled when compared to (a) (see pink box), increasing the distance between the nucleation sides of the instabilities (the lesser confined edges), domain walls can also emerge in these systems (red region). }
\label{SI_fig:1}
\end{figure}

 One could now argue that given a large enough system, and assuming that the initial instabilities are triggered at the lesser confined edges, lattices with a stronger coupling between layers, i.e., those with thicker bonds, might eventually also show domain walls.
To test this, we take the design of the cubic lattice indicated by the pink box in Fig.\,\ref{SI_fig:1}\textit{A} and extend it in one direction, creating a 10x5x5 lattice, as shown in Fig.\,\ref{SI_fig:1}\textit{B}.    
Whereas the front view does not clearly show the arising non-uniform pattern-formation during compression for this structure (left image), the side view clearly indicates the different orientation of the five layers on the left and on the right (green regions), separated again by a domain wall (red region). 
Apart from increasing the distance between the nucleation sides of the instabilities that then travel inwards through the lattice, one could also experiment by locally weakening the bonds connecting different layers by making them shorter, thinner or by using a material that is generally less stiff.  
The latter are future research directions that have not been explored yet.

\section*{II - The dynamic buckling equations }

In this section we provide a detailed derivation of the dynamic buckling equations for both uniaxial and isotropic compression.
We follow the line of thought as introduced on the main text, by modelling the interaction of the nodes through harmonic springs with different energies associated to them. 
Additionally we assume that all bonds dissipate energy depending on the angular velocities and neglect inertial effects. 
As mentioned in the main text we perform a coordinate transformation in line with the experimental observation on local buckling, assuming counter-rotation of neighbouring nodes that are in-plane.
Angles perpendicular to this plane will be aligned uniformly.
Transformation rules between both rotation basis are laid out below, as well as their affect on the overall structure of the dynamic buckling equations.

\subsection*{Uniaxial compression}

Starting the derivation by developing a discrete model, we consider the following energy contributions for bending, shearing, torsion, compression, as well as two Rayleigh dissipation functions, respectively:
\begin{equation*}
\begin{aligned}
E_b &= \frac{1}{2} k_b \sum_{i,j,k} \left[\theta_{i,j,k} + \theta_{i+1,j,k}\right]^2 +  \left[\theta_{i,j,k} + \theta_{i,j,k+1}\right]^2 +  \left[\phi_{i,j,k} + \phi_{i,j+1,k}\right]^2 +  \left[\phi_{i,j,k} + \phi_{i,j,k+1}\right]^2, \\
E_s &=  \frac{1}{2} k_s \left(\frac{L_{\mathrm{bond}}}{2}\right)^2 \sum_{i,j,k} \left[\theta_{i,j,k} - \theta_{i+1,j,k}\right]^2 +  \left[\theta_{i,j,k} - \theta_{i,j,k+1}\right]^2 +  \left[\phi_{i,j,k} - \phi_{i,j+1,k}\right]^2 +  \left[\phi_{i,j,k} - \phi_{i,j,k+1}\right]^2, \\ 
E_t &=  \frac{1}{2} k_t \sum_{i,j,k} \left[\theta_{i,j,k} - \theta_{i,j+1,k}\right]^2 +  \left[\phi_{i,j,k} - \phi_{i+1,j,k}\right]^2, \\
E_c &= -F L_{\mathrm{bond}} \sum_{i,j,k} \left[1 - \cos \theta_{i,j,k} \cos \phi_{i,j,k} \right],  \\
\mathcal{F}_{\theta} &=  \frac{1}{2}c_{\mathrm{disp}} \sum_{i,j,k} \dot{\theta}_{i,j,k}^2, \\
\mathcal{F}_{\phi} &=  \frac{1}{2}c_{\mathrm{disp}} \sum_{i,j,k} \dot{\phi}_{i,j,k}^2.
\end{aligned}
\end{equation*}
The coefficients are related to material properties as discussed in the main text. In short, $k_b = E_{\mathrm{bond}}I_{\mathrm{bond}}/L_{\mathrm{bond}}$, $k_s = \kappa G_{\mathrm{bond}}A_{\mathrm{bond}}/L_{\mathrm{bond}} = \frac{1-\zeta}{\zeta} G_{\mathrm{bond}}A_{\mathrm{bond}} R_{\mathrm{bond}}/L_{\mathrm{bond}}^2$, $k_t = G_{\mathrm{bond}}J_{\mathrm{bond}}/_{\mathrm{bond}}L$, $F = N^2 E_{\mathrm{eff}}A_{\mathrm{bond}}\epsilon = N^2 k_c L_{\mathrm{bond}}^2 \epsilon$, with $k_c = E_{\mathrm{eff}}A_{\mathrm{bond}}/L_{\mathrm{bond}}$. 
The area and polar moments in inertia are $I_{\mathrm{bond}} = \pi R_{\mathrm{bond}}^4/4$ and $J_{\mathrm{bond}} = \pi R_{\mathrm{bond}}^4/2$.

We now consider potential energy $U = E_b + E_s + E_t + E_c$ and evaluate the simplified Lagrangian formulations $\partial_{\theta_{i,j,k}} (E_b + E_s + E_c + E_t)  + \partial_{\dot{\theta}_{i,j,k}} \mathcal{F}_{\theta} = 0$, as well as $\partial_{\phi_{i,j,k}} (E_b + E_s + E_c + E_t)  + \partial_{\dot{\phi}_{i,j,k}} \mathcal{F}_{\phi} = 0$.
The derivatives read:
\begin{equation*}
\begin{aligned}
\partial_{\theta_{i,j,k}} U &= k_b \left[4 \theta_{i,j,k} + \theta_{i+1,j,k} + \theta_{i-1,j,k} + \theta_{i,j,k+1} + \theta_{i,j,k-1} \right] \\
& + k_s \left(L_{\mathrm{bond}}/2\right)^2  \left[4 \theta_{i,j,k} - \theta_{i+1,j,k} - \theta_{i-1,j,k} - \theta_{i,j,k+1} - \theta_{i,j,k-1} \right] \\
& + k_t \left[2 \theta_{i,j,k} - \theta_{i,j+1,k} - \theta_{i,j-1,k} \right] \\
& - F L_{\mathrm{bond}} \sin \theta_{i,j,k} \cos \phi_{i,j,k},
\end{aligned}
\end{equation*}
\begin{equation*}
\begin{aligned}
\partial_{\phi_{i,j,k}} U &= k_b \left[4 \phi_{i,j,k} + \phi_{i,j+1,k} + \phi_{i,j-1,k} + \phi_{i,j,k+1} + \phi_{i,j,k-1} \right] \\
& + k_s\left(L_{\mathrm{bond}}/2\right)^2  \left[4 \phi_{i,j,k} - \phi_{i,j+1,k} - \phi_{i,j-1,k} - \phi_{i,j,k+1} - \phi_{i,j,k-1} \right] \\
& + k_t \left[2 \phi_{i,j,k} - \phi_{i+1,j,k} - \phi_{i-1,j,k} \right] \\
& - F L_{\mathrm{bond}} \sin \phi_{i,j,k} \cos \theta_{i,j,k},
\end{aligned}
\end{equation*}
Eventually, we obtain two discrete equations as follows:
\begin{equation*}
\begin{aligned}
c_{\mathrm{disp}}\dot{\theta}_{i,j,k}&= -k_b \left[4 \theta_{i,j,k} + \theta_{i+1,j,k} + \theta_{i-1,j,k} + \theta_{i,j,k+1} + \theta_{i,j,k-1} \right] \\
& - k_s\left(L_{\mathrm{bond}}/2\right)^2  \left[4 \theta_{i,j,k} - \theta_{i+1,j,k} - \theta_{i-1,j,k} - \theta_{i,j,k+1} - \theta_{i,j,k-1} \right] \\
& - k_t \left[2 \theta_{i,j,k} - \theta_{i,j+1,k} - \theta_{i,j-1,k} \right] \\
& + F L_{\mathrm{bond}} \sin \theta_{i,j,k} \cos \phi_{i,j,k},
\end{aligned}
\end{equation*}
\begin{equation*}
\begin{aligned}
c_{\mathrm{disp}}\dot{\phi}_{i,j,k}&= -k_b \left[4 \phi_{i,j,k} + \phi_{i,j+1,k} + \phi_{i,j-1,k} + \phi_{i,j,k+1} + \phi_{i,j,k-1} \right] \\
& - k_s\left(L_{\mathrm{bond}}/2\right)^2  \left[4 \phi_{i,j,k} - \phi_{i,j+1,k} - \phi_{i,j-1,k} - \phi_{i,j,k+1} - \phi_{i,j,k-1} \right] \\
& - k_t \left[2 \phi_{i,j,k} - \phi_{i+1,j,k} - \phi_{i-1,j,k} \right] \\
& + F L_{\mathrm{bond}} \sin \phi_{i,j,k} \cos \theta_{i,j,k}.
\end{aligned}
\end{equation*}
Next, we approximate the discrete differences by continuous functions through Taylor expansions, approaching the continuum limit. 
As a consequence:
\begin{equation*}
\begin{aligned}
&\theta_{i+1,j,k} + \theta_{i-1,j,k} - 2 \theta_{i,j,k} \rightarrow L_{\mathrm{bond}}^2 \frac{\partial^2 \theta}{\partial x^2}, \, \theta_{i,j+,k} + \theta_{i,j-1,k} - 2 \theta_{i,j,k} \rightarrow L_{\mathrm{bond}}^2 \frac{\partial^2 \theta}{\partial y^2}, \, \theta_{i,j,k+1} + \theta_{i,j,k-1} - 2 \theta_{i,j,k} \rightarrow L_{\mathrm{bond}}^2 \frac{\partial^2 \theta}{\partial z^2}, \\
&\phi_{i+1,j,k} + \phi_{i-1,j,k} - 2 \phi_{i,j,k} \rightarrow L_{\mathrm{bond}}^2 \frac{\partial^2 \phi}{\partial x^2}, \, \phi_{i,j+1,k} + \phi_{i,j-1,k} - 2 \phi_{i,j,k} \rightarrow L_{\mathrm{bond}}^2 \frac{\partial^2 \phi}{\partial y^2}, \, \phi_{i,j,k+1} + \phi_{i,j,k-1} - 2 \phi_{i,j,k} \rightarrow L_{\mathrm{bond}}^2 \frac{\partial^2 \phi}{\partial z^2}.
\end{aligned}
\end{equation*}
Additionally, assuming small angles we approximate that $\sin \theta \cos \phi \approx \left(\theta - \theta^3/6 \right)\left(1 - \phi^2/2 \right) \approx \theta - \theta^3/6 - \theta \phi^2/2 $ and $\sin \phi \cos \theta \approx \left(\phi - \phi^3/6 \right)\left(1 - \theta^2/2 \right) \approx \phi - \phi^3/6 - \phi \theta^2/2 $.

Combining and rewriting yields
\begin{equation*}
\begin{aligned}
c_{\mathrm{disp}} \dot{\theta} &= \left(k_s L_{\mathrm{bond}}^2/4 - k_b \right)L_{\mathrm{bond}}^2 \nabla^2_{xz} \theta + k_t L_{\mathrm{bond}}^2 \frac{\partial^2 \theta}{\partial y^2} + F L_{\mathrm{bond}} \theta - 8 k_b \theta -  \frac{F L_{\mathrm{bond}}}{6} \theta^3 - \frac{F L_{\mathrm{bond}}}{2}\theta \phi^2, \\
c_{\mathrm{disp}} \dot{\phi} &= \left(k_s L_{\mathrm{bond}}^2/4 - k_b \right)L_{\mathrm{bond}}^2 \nabla^2_{yz} \phi + k_t L_{\mathrm{bond}}^2 \frac{\partial^2 \phi}{\partial x^2} + F L_{\mathrm{bond}} \phi - 8 k_b \phi -  \frac{F L_{\mathrm{bond}}}{6} \phi^3 - \frac{F L_{\mathrm{bond}}}{2}\phi \theta^2.
\end{aligned}
\end{equation*}
Finally, we can obtain Eq.\,3\,\&\,4 from the main text:
\begin{equation*}
\begin{aligned}
\dot{\theta} = & D_{xz} \nabla_{xz}^2 \theta + D_{y}  \frac{\partial^2 \theta}{\partial y^2} + f(\theta,\phi), \\
\dot{\phi}=  & D_{yz} \nabla_{yz}^2 \phi + D_{x}  \frac{\partial^2 \phi}{\partial x^2} + g(\theta,\phi),
\end{aligned}
\end{equation*}
with $D_{xz} = D_{yz} = \left(k_s L_{\mathrm{bond}}^2/4 - k_b \right)L_{\mathrm{bond}}^2 / c_{\mathrm{disp}}$ and $D_y = D_x = k_t L_{\mathrm{bond}}^2 / c_{\mathrm{disp}}$ the diffusivity constants in the xz- and yz-planes, and y- and x-directions, respectively. 
Also, $\nabla_{xz}^2 = \partial^2/\partial x^2 + \partial^2 / \partial z^2 $ and $\nabla_{yz}^2 = \partial^2/\partial y^2 + \partial^2 / \partial z^2 $, and the forcing terms read:
\begin{equation*}
\begin{aligned}
f(\theta,\phi) = & \mathcal{C}_{\theta} \left[1  - \frac{8 k_b}{F L_{\mathrm{bond}}}  - \frac{\theta^2}{6} - \frac{\phi^2}{2} \right] \theta,  \\
g(\theta,\phi) = & \mathcal{C}_{\phi}  \left[1  - \frac{8 k_b}{F L_{\mathrm{bond}}}  - \frac{\phi^2}{6} - \frac{\theta^2}{2} \right] \phi,  
\end{aligned}
\end{equation*}
where $\mathcal{C}_{\theta} = \mathcal{C}_{\phi} = F L_{\mathrm{bond}} / c_{\mathrm{disp}}$.

\subsection*{Isotropic compression}

We can repeat the derivation considering also rotation in the xy-plane, which we define as $\psi$.
Additionally, we can apply isotropic compression with forces in the three principle directions, namely $F_x$, $F_y$, and $F_z$ for a more generalized description. 
Without providing the derivation, the obtained system of equations becomes:
\begin{equation*}
\begin{aligned}
\dot{\theta} = & D_{xz} \nabla_{xz}^2 \theta + D_{y}  \frac{\partial^2 \theta}{\partial y^2} + f(\theta,\phi,\psi), \\
\dot{\phi}=  & D_{yz} \nabla_{yz}^2 \phi + D_{x}  \frac{\partial^2 \phi}{\partial x^2} + g(\theta,\phi,\psi),\\
\dot{\psi}=  & D_{xy} \nabla_{xy}^2 \psi + D_{z}  \frac{\partial^2 \psi}{\partial z^2} + h(\theta,\phi,\psi),
\end{aligned}
\end{equation*}
with $D_{xz} = D_{yz} = D_{xy} = \left(k_s L_{\mathrm{bond}}^2/4 - k_b \right)L_{\mathrm{bond}}^2 / c_{\mathrm{disp}}$ and $D_y = D_x = D_z = k_t L_{\mathrm{bond}}^2 / c_{\mathrm{disp}}$ the diffusivity constants in the xz-, yz-, and xy-planes, and y-, x-, and z-directions, respectively. 
Also, $\nabla_{xz}^2 = \partial^2/\partial x^2 + \partial^2 / \partial z^2 $, $\nabla_{yz}^2 = \partial^2/\partial y^2 + \partial^2 / \partial z^2 $, and $\nabla_{xy}^2 = \partial^2/\partial x^2 + \partial^2 / \partial y^2 $. 
The forcing terms read:
\begin{equation*}
\begin{aligned}
f(\theta,\phi,\psi) = & \mathcal{C}_{\theta} \left[1  - \frac{8 k_b}{\left(F_x + F_z \right)L_{\mathrm{bond}} }  - \frac{\theta^2}{6}  \right] \theta - \frac{L_{\mathrm{bond}}}{2}\left[F_x \psi^2 + F_z \phi^2 \right]\theta,  \\
g(\theta,\phi,\psi) = & \mathcal{C}_{\phi} \left[1  - \frac{8 k_b}{\left(F_y + F_z \right)L_{\mathrm{bond}} }  - \frac{\phi^2}{6}  \right] \phi - \frac{L_{\mathrm{bond}}}{2}\left[F_y \psi^2 + F_z \theta^2 \right]\phi,  \\
h(\theta,\phi,\psi) = & \mathcal{C}_{\psi} \left[1  - \frac{8 k_b}{\left(F_x + F_y \right)L_{\mathrm{bond}} }  - \frac{\phi^2}{6}  \right] \psi - \frac{L_{\mathrm{bond}}}{2}\left[F_y \phi^2 + F_x \theta^2 \right]\psi,  
\end{aligned}
\end{equation*}
where $\mathcal{C}_{\theta} = \left(F_x + F_z \right) L_{\mathrm{bond}} / c_{\mathrm{disp}}$, $\mathcal{C}_{\phi} = \left(F_y + F_z \right) L_{\mathrm{bond}} / c_{\mathrm{disp}}$, and $\mathcal{C}_{\psi} = \left(F_x + F_y \right) L_{\mathrm{bond}} / c_{\mathrm{disp}}$.
Taking $F_x = F_y = \psi = 0$ yields the uniaxial compression case for $F_z = F$.

\subsection*{Derivation in the co-rotating reference frame}

As mentioned above, the derivation is performed in the counter-rotating reference frame with counter-rotating neighbours in-plane and aligned rotation in the perpendicular direction.
The following transformation rules between 'global' co-rotation ($\theta^{\star}$) and 'local' counter-rotation ($\theta$) hold:
\begin{equation*}
\theta_{i,j,k}^{\star} = \left( -1 \right)^{i+k}\theta_{i,j,k}, \quad \phi_{i,j,k}^{\star} = \left( -1 \right)^{j+k}\phi_{i,j,k}, \quad \psi_{i,j,k}^{\star} = \left( -1 \right)^{i+j}\psi_{i,j,k}.
\end{equation*}
If we change to the co-rotating frame of reference (which more resembles the nodal rotation corresponding to global buckling, see Fig.\,1\textit{C} in the main text), the above equations become:
\begin{equation*}
\begin{aligned}
\dot{\theta}^{\star} = & D_{xz}^{\star} \nabla_{xz}^2 \theta^{\star} + D_{y}  \frac{\partial^2 \theta^{\star}}{\partial y^2} + f(\theta^{\star},\phi^{\star},\psi^{\star}), \\
\dot{\phi}^{\star}=  & D_{yz}^{\star} \nabla_{yz}^2 \phi^{\star} + D_{x}  \frac{\partial^2 \phi^{\star}}{\partial x^2} + g(\theta^{\star},\phi^{\star},\psi^{\star}),\\
\dot{\psi}^{\star}=  & D_{xy}^{\star} \nabla_{xy}^2 \psi^{\star} + D_{z}  \frac{\partial^2 \psi^{\star}}{\partial z^2} + h(\theta,\phi^{\star},\psi^{\star}),
\end{aligned}
\end{equation*}
with $D_{xz}^{\star} = D_{yz}^{\star} = D_{xy}^{\star} = \left(k_b - k_s L_{\mathrm{bond}}^2/4 \right)L_{\mathrm{bond}}^2 / c_{\mathrm{disp}}$.
The forcing terms read:
\begin{equation*}
\begin{aligned}
f(\theta^{\star},\phi^{\star},\psi^{\star}) = & \mathcal{C}_{\theta} \left[1  - \frac{8 k_s L_{\mathrm{bond}}^2}{\left(F_x + F_z \right)L_{\mathrm{bond}} }  - \frac{{\theta^{\star}}^2}{6}  \right] \theta^{\star} - \frac{L_{\mathrm{bond}}}{2}\left[F_x {\psi^{\star}}^2 + F_z {\phi^{\star}}^2 \right]\theta^{\star},  \\
g(\theta^{\star},\phi^{\star},\psi^{\star}) = & \mathcal{C}_{\phi} \left[1  - \frac{8 k_s L_{\mathrm{bond}}^2}{\left(F_y + F_z \right)L_{\mathrm{bond}} }  - \frac{{\phi^{\star}}^2}{6}  \right] \phi^{\star} - \frac{L_{\mathrm{bond}}}{2}\left[F_y {\psi^{\star}}^2 + F_z {\theta^{\star}}^2 \right]\phi^{\star},  \\
h(\theta^{\star},\phi^{\star},\psi^{\star}) = & \mathcal{C}_{\psi} \left[1  - \frac{8 k_s L_{\mathrm{bond}}^2}{\left(F_x + F_y \right)L_{\mathrm{bond}} }  - \frac{{\phi^{\star}}^2}{6}  \right] \psi^{\star} - \frac{L_{\mathrm{bond}}}{2}\left[F_y {\phi^{\star}}^2 + F_x {\theta^{\star}}^2 \right]\psi^{\star}.
\end{aligned}
\end{equation*}
In the main text we argue that we transition from global to local buckling when $k_b/(k_s L_{\mathrm{bond}}^2) = 1$, which is a direct consequence for the case when $D_{\theta} = D_{\theta}^{\star}$.
This is also criteria that is the bases of the derivation of Eq.\,1 in the main text.

\subsection*{Buckling plane selection}
Here we analytically show that the selection of a buckling plane will persist, \textit{i.e.}, $\theta \neq 0$ and $\phi = 0$ or vice versa, for uniaxial compression.
As stated in the main text, we assume a potential energy $V(\theta,\phi)$ for which $\partial V(\theta,\phi)/\partial \theta = - c_{\mathrm{disp}} f(\theta,\phi) = 0$ and $\partial V(\theta,\phi)/\partial \phi = - c_{\mathrm{disp}} g(\theta,\phi) = 0$ need to be satisfied. 
If both variables are non-zero, we can show they must be equal, hence $\theta^2 = \phi^2 = \xi^2$.
Evaluating the terms in brackets of $f(\theta,\phi)$ or $g(\theta,\phi)$ then yields
\begin{equation*}
\theta^2 = \phi^2 = \xi^2 = \frac{3 \left(1 - \frac{8 k_b}{F L_{\mathrm{bond}}} \right)}{2}.
\end{equation*}
Integration of $f(\theta,\phi)$ and $g(\theta,\phi)$ results in a potential energy as
\begin{equation}
V(\theta,\phi) =  - c_{\mathrm{disp}}\mathcal{C} \left[ \left( 1 -\frac{8 k_b}{F L_{\mathrm{bond}}} \right)\frac{ \theta^2 + \phi^2 }{2} - \frac{\theta^4 + \phi^4}{24} - \frac{\theta^2 \phi^2}{4} \right],
\end{equation}
with $\mathcal{C}=\mathcal{C}_{\theta}=\mathcal{C}_{\phi}$.  The Hessian of this function reads
\begin{equation*}
H \propto \begin{pmatrix}
\frac{8 k_b}{F L_{\mathrm{bond}}} -1 + \frac{\theta^2}{2} + \frac{\phi^2}{2} & \theta \phi \\
 \theta \phi  & \frac{8 k_b}{F L_{\mathrm{bond}}} -1 + \frac{\phi^2}{2} + \frac{\theta^2}{2}
\end{pmatrix}.
\end{equation*}
Let $\theta^2 = \phi^2 = \xi^2$, then
\begin{equation*}
\begin{aligned}
a & = \frac{8k_b}{F L_{\mathrm{bond}}}  - 1 + \xi^2 = \frac{1}{2} \left[ 1 - \frac{8k_b}{F L_{\mathrm{bond}}} \right], \\
b &=  \theta \phi  = \pm\xi^2 = \pm\frac{3}{2} \left[ 1 - \frac{8k_b}{F L_{\mathrm{bond}}} \right].
\end{aligned}
\end{equation*}
Hence,
\begin{equation*}
\det(H) \propto a^2 - b^2 = \frac{1}{4}\left[ 1 - \frac{8k_b}{F L_{\mathrm{bond}}} \right]^2 - \frac{9}{4}\left[ 1 - \frac{8k_b}{F L_{\mathrm{bond}}} \right]^2 = -2 \left[ 1 - \frac{8k_b}{F L_{\mathrm{bond}}} \right]^2 < 0.
\end{equation*}
Since $\det(H)<0$ this resembles a saddle point, which is unstable. 
For the selection of a buckling plane (i.e., $\theta \neq 0$, $\phi = 0$ or vice versa) the Hessian is diagonal with positive diagonal elements.
This represents a stable minimum and agrees with our experimental observations (see Fig.\,1\textit{B}).

\section*{III - Strain energy dissipation dependence on lattice structure }

Here we look into the energy that is being dissipated by the three-dimensional simple cubic lattices during one loading cycle.
It allows us to quantify how sensitive the strain energy dissipation is with respect to the architecture of the lattice structure.
Since the sizes of the cubic lattices are not hugely different and we always use the same crosshead speed during compression, variations in strain rate $\dot{\epsilon}$ are assumed to be negligible.
The affect of strain rate on the loss modulus (or here the energy dissipated during one loading cycle) is beyond the scope of the current study.

\begin{figure}[t!]
  \centerline{\includegraphics[width=\textwidth]{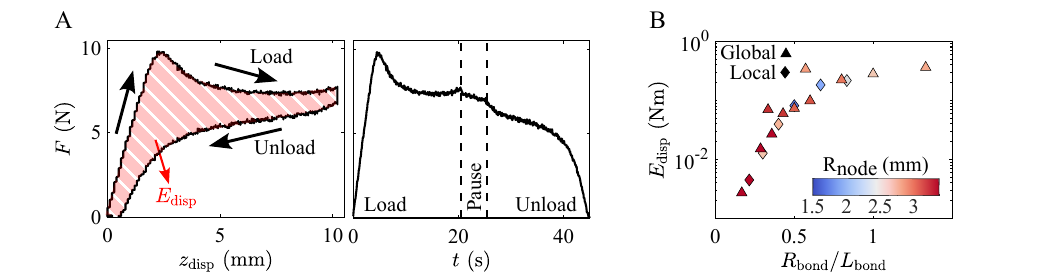}}
\caption{\textbf{Strain energy dissipation.}
(\textit{A}) Force as a function of compression $z_{\mathrm{disp}}$ (left) and time (right) for a simple cubic lattice with $N = 5$, $R_{\mathrm{node}} = \SI{2}{\mm}$, $R_{\mathrm{bond}} = \SI{1.25}{\mm}$, and $L = \SI{30}{\mm}$.
Between loading and unloading we pause for $\SI{5}{\s}$. 
The red shaded area with the force versus displacement loop defines the energy dissipated $E_{\mathrm{disp}}$ by the cubic lattice during one loading cycle.
(\textit{B}) Averaged dissipated energy $E_{\mathrm{disp}}$ as a function of bond radius over bond length for all designs over three loading cycles.}
\label{SI_fig:2}
\end{figure}

The  loading cycle consists out of a loading stage up to 0.3 strain with crosshead speed $v = \SI{0.5}{\mm \per \s}$, followed by a pause of $\SI{5}{\s}$, and subsequent unloading at the same speed.
A typical force versus displacement curve is shown in Fig.\,\ref{SI_fig:2}\textit{A}, as well as force versus time.
After the linear response during loading the lattice buckles and weakens. 
During the pause, some energy is dissipated, reflecting the viscoelasticity in the 3D-printed material, and even more so during unloading.
The mechanical work associated with the force-displacement cycle is calculated by integrating the force over displacement,
\begin{equation*}
    E = \int_{\mathcal{\delta}} F(z)\,\mathrm{d}z,
\end{equation*}
where $\mathcal{\delta}$ denotes the relevant loading or unloading path. During loading, the work input is
\begin{equation*}
    E_{\mathrm{load}} = \int_{0}^{z_{\max}} F_{\mathrm{load}}(z)\,\mathrm{d}z,
\end{equation*}
while during unloading it is given by
\begin{equation*}
    E_{\mathrm{unload}} = \int_{z_{\max}}^{0} F_{\mathrm{unload}}(z)\,\mathrm{d}z.
\end{equation*}
By construction, $E_{\mathrm{load}}>0$ and $E_{\mathrm{unload}}<0$. 
The dissipated energy, corresponding to the area enclosed by the force-displacement loop, is then $E_{\mathrm{disp}} = E_{\mathrm{load}} - \vert E_{\mathrm{unload}}\vert$.
We define the dissipation ratio as $\eta = E_{\mathrm{disp}} /E_{\mathrm{load}} = 1 - \vert E_{\mathrm{unload}} \vert /E_{\mathrm{load}}$ (see Fig.\,2\textit{F} in the main text).

\section*{IV - Loading rate dependence on mechanical response, strain energy dissipation and nodal rotation dynamics }
\begin{figure}[h!]
  \centerline{\includegraphics[width=\columnwidth]{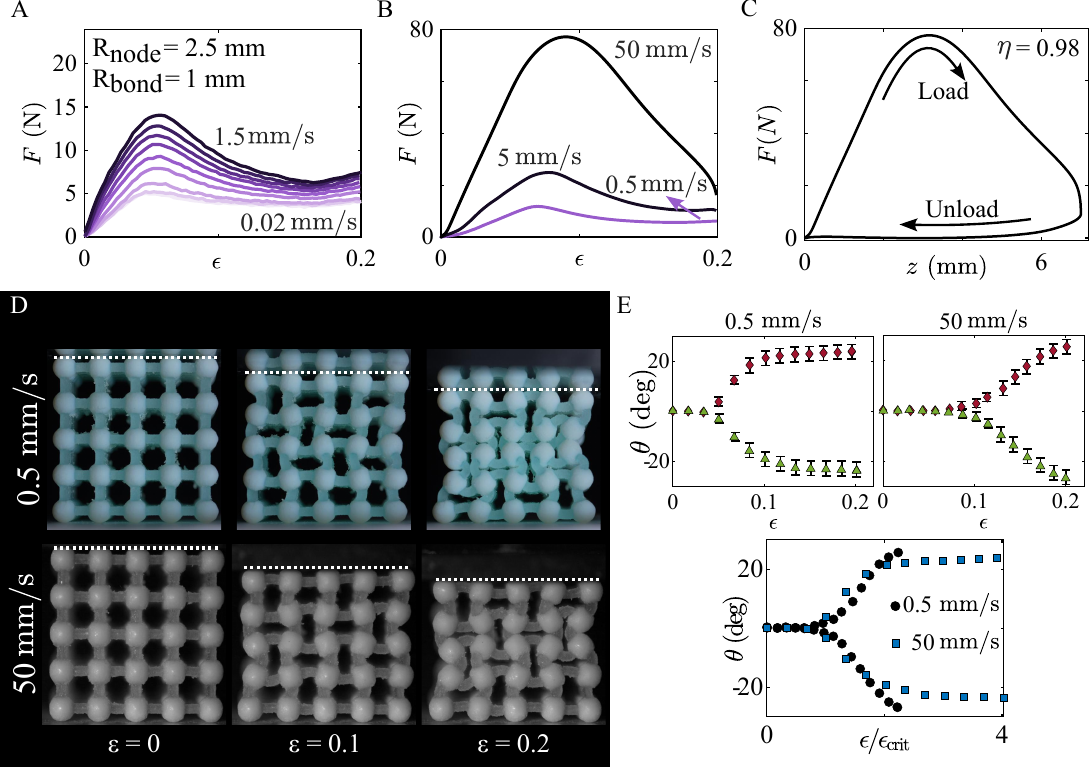}}
\caption{ (\textit{A-B}) Force strain relationships of a typical structure with $N = 5$, $R_{\mathrm{bond}} = \SI{1}{\mm}$, $R_{\mathrm{node}} = \SI{2.5}{\mm}$ and $L_{\mathrm{bond}} = \SI{2.5}{\mm}$ at different crosshead velocities. Curves are averages of three repetitions. 
(\textit{C}) Force versus displacement during a loading and unloading cycle of the same structure at $v_{\mathrm{crosshead}} = \SI{50}{\mm \per \s}$. The strain energy dissipation ratio is determined to be $\eta =0.98$.
(\textit{D}) Experimental snapshots of the lattice structure during slow (top) and fast (bottom) compression.
(\textit{E}) Nodal rotations of the center nine nodes for the slow (top left) and fast (top right) compression. Bottom: Node rotations as a function of $\epsilon/\epsilon_{\mathrm{crit}}$.
}
\label{SI_fig:3}
\end{figure}

The mechanical response of a typical structure with $N = 5$, $R_{\mathrm{bond}} = \SI{1}{\mm}$, $R_{\mathrm{node}} = \SI{2.5}{\mm}$ and $L_{\mathrm{bond}} = \SI{2.5}{\mm}$ is shown here in Fig.\,\ref{SI_fig:3}\textit{A}\&\textit{B} for crosshead velocities ranging from $\SI{0.02}{\mm \per \s}$ to $\SI{50}{\mm \per \s}$. 
Whereas the mechanical response seems to converge at low straining velocities, the larger the velocity becomes, the stiffer the structure gets. This is expected due to the viscoelastic nature of the polymers. 
Apart from the change in stiffness, also the critical strain at which buckling occurs shifts towards larger values as the compression speed is increased. 

When unloading the structure after loading at $\SI{50}{\mm \per \second}$ with the same velocity, the structure has dissipated some energy (see Fig.\,\ref{SI_fig:3}\textit{C}). 
When determining the strain dissipation ratio as discussed in the main text we find a value of $\eta = 0.98$.
This is significantly larger as compared to loading/unloading at a lower velocity (see Fig.\,2\textit{F} in the main text). 

Although the mechanical response and the energy dissipation are highly dependent on the loading-rate, the most important feature for our study, namely the dynamics of the node rotations, remain unchanged upon proper normalisation. 
We show this in Fig.\,\ref{SI_fig:3}\textit{D}\&\textit{E}.
Fig.\,\ref{SI_fig:3}\textit{D} shows experimental snapshots of the compression of the same lattice structure at different loading-rates at the same values of applied strain, exhibiting striking similarities.  
To quantify this, we look into the node rotations of the center nine nodes (as also done in Fig.\,1\textit{C} of the main text) in Fig.\,\ref{SI_fig:3}\textit{E}. 
The upper plots show $\theta$ as a function of strain for the slow (left) and fast (right) compression. 
The lower plots shows their collapse when normalising strain with the corresponding value of the critical strain at which buckling occurs.

\section*{V - Numerical implementation }

In order to evaluate Eqs.\,3\,\&\,5 of the main text  numerically we non-dimensionalize the system of equations.
To repeat, we have
\begin{equation*}
\begin{aligned}
\partial_t \theta(\vec{x},t) = & D_{xz} \left(\partial_x^2 + \partial_z^2 \right) \theta(\vec{x},t) + D_{y}  \partial_y^2 \theta(\vec{x},t) - \mathcal{R} \epsilon_{\mathrm{crit}} \theta(\vec{x},t) + \mathcal{R} \epsilon \theta(\vec{x},t) + \mathcal{R} \epsilon \frac{\theta(\vec{x},t)^3}{6} + \mathcal{R} \epsilon \frac{\theta(\vec{x},t) \phi(\vec{x},t)^2}{2} , \\
\partial_t \phi(\vec{x},t) =  & D_{yz} \left(\partial_y^2 + \partial_z^2 \right) \phi(\vec{x},t) + D_{x}  \partial_x^2 \phi(\vec{x},t)  - \mathcal{R} \epsilon_{\mathrm{crit}}\phi(\vec{x},t) + \mathcal{R} \epsilon \phi(\vec{x},t) + \mathcal{R} \epsilon \frac{\phi(\vec{x},t)^3}{6} + \mathcal{R} \epsilon \frac{\phi(\vec{x},t) \theta(\vec{x},t)^2}{2},
\end{aligned}
\end{equation*}
with $D_{xz} = D_{yz} = \left(k_s L_{\mathrm{bond}}^2/4 - k_b \right) L_{\mathrm{bond}}^2 / c_{\mathrm{disp}}$, $D_y = D_x = k_t  L_{\mathrm{bond}}^2 / c_{\mathrm{disp}}$, and $\mathcal{R} = N^2 k_c  L_{\mathrm{bond}}^2/ c_{\mathrm{disp}}$.
We then consider the following substitutions:
\begin{equation*}
\tilde{t} = \frac{t}{\tau_{\mathrm{diff}}}, \quad \tilde{\vec{x}} = \frac{\vec{x}}{L_{\mathrm{bond}}}, \quad \tilde{\theta} = \frac{\theta}{\theta_{\mathrm{max}}}, \quad  \tilde{\phi} = \frac{\phi}{\phi_{\mathrm{max}}},
\end{equation*}
where tildes denote dimensionless quantities.
$\tau_{\mathrm{diff}}$ is the experimentally observed diffusion time scale, e.g., the time it takes in the experiments for diffusion to have propagated throughout the whole system. 
The latter is approximately $\SI{50}{\second}$ for the case shown in Fig.\,5 in the main text.
 $\theta_{\mathrm{max}} = \phi_{\mathrm{max}} = \SI{40}{\degree} = \frac{2 \pi}{9} \mathrm{rad}$ is the experimentally observed maximum node rotation during compression and also used as upper constraint in the simulations.
Substitution then yields
\begin{equation*}
\begin{aligned}
\partial_{\tilde{t}} \tilde{\theta} = & \frac{D_{xz} \tau_{\mathrm{diff}}}{L_{\mathrm{bond}}^2} \left(\partial_{\tilde{x}}^2 + \partial_{\tilde{z}}^2 \right) \tilde{\theta} + \frac{D_{y}\tau_{\mathrm{diff}}}{L_{\mathrm{bond}}^2} \partial_{\tilde{y}}^2 \tilde{\theta} - \mathcal{R} \tau_{\mathrm{diff}} \epsilon_{\mathrm{crit}} \tilde{\theta} + \mathcal{R} \tau_{\mathrm{diff}} \epsilon \tilde{\theta} + \mathcal{R} \tau_{\mathrm{diff}} \epsilon \frac{\tilde{\theta}^3}{6\theta_{\mathrm{max}}^2} + \mathcal{R} \tau_{\mathrm{diff}} \epsilon \frac{\tilde{\theta} \tilde{\phi}^2}{2 \phi_{\mathrm{max}}^2} , \\
\partial_{\tilde{t}} \tilde{\phi} =  & \frac{D_{yz} \tau_{\mathrm{diff}}}{L_{\mathrm{bond}}^2} \left(\partial_{\tilde{y}}^2 + \partial_{\tilde{z}}^2 \right) \tilde{\phi} + \frac{D_{x} \tau_{\mathrm{diff}}}{L_{\mathrm{bond}}^2} \partial_{\tilde{x}}^2 \tilde{\phi}  - \mathcal{R} \tau_{\mathrm{diff}} \epsilon_{\mathrm{crit}} \tilde{\phi} + \mathcal{R} \tau_{\mathrm{diff}} \epsilon \tilde{\phi} + \mathcal{R} \tau_{\mathrm{diff}} \epsilon \frac{\tilde{\phi}^3}{6 \phi_{\mathrm{max}}^2} + \mathcal{R} \tau_{\mathrm{diff}} \epsilon \frac{\tilde{\phi} \tilde{\theta}^2}{2 \theta_{\mathrm{max}}^2}.
\end{aligned}
\end{equation*}
We obtain dimensionless groups 
\begin{equation*}
\tilde{D}_{xz} = \frac{D_{xz} \tau_{\mathrm{diff}}}{L_{\mathrm{bond}}^2}, \quad \tilde{D}_{yz} = \frac{D_{yz} \tau_{\mathrm{diff}}}{L_{\mathrm{bond}}^2}, \quad \tilde{D}_y = \frac{D_{y}\tau_{\mathrm{diff}}}{L_{\mathrm{bond}}^2}, \quad \tilde{D}_x = \frac{D_{x}\tau_{\mathrm{diff}}}{L_{\mathrm{bond}}^2}, \quad \tilde{\mathcal{R}} = \mathcal{R} \tau_{\mathrm{diff}}.
\end{equation*}
We numerically solve the dimensionless equations by standard finite, first order difference discretization of the second-order derivatives on a three-dimensional grid with $L_x = L_y = L_z = 50$ points and lattice spacing $dx = dy = dz = 1$. 
Time integration is performed using a forward (explicit) scheme and the time-step $dt \ll 1$ is kept small to ensure stability. 

As initial conditions, we assume random initial perturbations of the order of 1\% for both angles, as well as stronger Gaussian initial perturbations in line with experimental observations regarding their origin (the least confined edges); see Fig.\,4\textit{A}\&\textit{B} of the main text, respectivly.
We also consider $\epsilon = \epsilon_0 + \dot{\epsilon} \tilde{t}$ - in line with the strain driven experiments - with an initial value for $\epsilon_0$ that can be lower, equal to or larger than $\epsilon_{\mathrm{crit}}$, where typically $\epsilon_{\mathrm{crit}} = 0.05$.
The strain rate can be chosen such to match experiments, with $\dot{\epsilon} = \SI{3e-3}{\per \second}$ as typical value for the case corresponding to Fig.\,5 of the main text, resulting in $\tilde{\dot{\epsilon}} = \dot{\epsilon} \tau_{\mathrm{diff}}$ in the dimensionless simulations.

\begin{figure}[b!]
  \centerline{\includegraphics[width=\columnwidth]{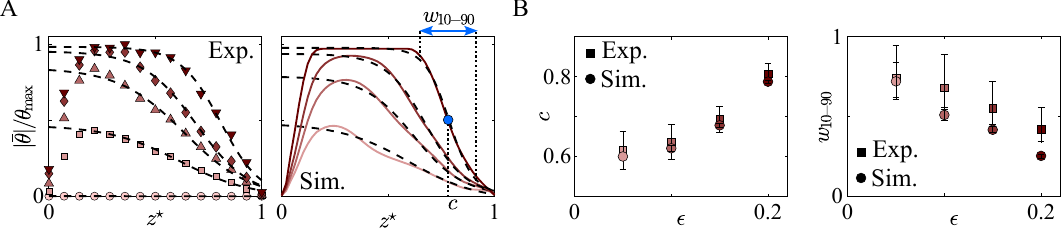}}
\caption{ (\textit{A}) Normalized averages of the rotation angles as a function of rescaled vertical coordinate $z^{\star}$ at different instances in time (or strain), for crosshead speed $\SI{0.5}{\mm \per \s}$.
The dashed lines correspond to fitted hyperbolic tangents.
(\textit{B} Obtained values of the center $c$ and its 10\%-to-90\% width $w_{10-90}$ for the experiment (squares) and simulation (circles). The slope of the left plot is representative for the front velocity since $\epsilon = \epsilon_0 + \dot{\epsilon} \tilde{t}$.) 
}
\label{SI_fig:4}
\end{figure}

\subsection*{Quantitive comparison to experiments.}
In Fig.\,5\textit{C} in the main text we have compared the evolution of the buckling front between 2D experiments and simulations. 
Here we provide additional details regarding their quantitive agreement. 

Fig.\,\ref{SI_fig:4}\textit{A} corresponds to Fig.\,5\textit{C} of the main text showing the mean absolute angle normalised by $\theta_{\mathrm{max}} = \SI{40}{\degree} = \frac{2 \pi}{9} \mathrm{rad}$ as a function of dimensionless and rescaled vertical postion $z^{\star} = 1 - \frac{z}{L(1-\epsilon)}$.
Different symbols (experimet, left) and colours (simulation, right) correspond to instances in time at equal strain, with $\epsilon = [0.05, 0.1, 0.15, 0.2]$, respectively. 
The circles in the left plot of Fig.\,\ref{SI_fig:4}\textit{A} correspond to $\epsilon = 0$.

For the simulations the system of equations simplifies due to the 2D approximation ($D_y = D_x = \phi = 0$), resulting in
\begin{equation*}
\begin{aligned}
\partial_{\tilde{t}} \tilde{\theta} = & \frac{D_{xz} \tau_{\mathrm{diff}}}{L_{\mathrm{bond}}^2} \left(\partial_{\tilde{x}}^2 + \partial_{\tilde{z}}^2 \right) \tilde{\theta}  - \mathcal{R} \tau_{\mathrm{diff}} \epsilon_{\mathrm{crit}} \tilde{\theta} + \mathcal{R} \tau_{\mathrm{diff}} \epsilon \tilde{\theta} + \mathcal{R} \tau_{\mathrm{diff}} \epsilon \frac{\tilde{\theta}^3}{6 \theta_{\mathrm{max}}^2}.
\end{aligned}
\end{equation*}
As an initial condition we consider the experimental snapshot at $\epsilon = 0.05$.
We now can match the representative curves of the evolving buckling fronts considering that the front takes the shape of a hyperbolic tangent (show in Suppl.\,Mat.\,Sec.\,XX) to determine the approximate values of $\tilde{D}_{xz} = \frac{D_{xz} \tau_{\mathrm{diff}}}{L_{\mathrm{bond}}^2}$ and $\tilde{\mathcal{R}} = \mathcal{R} \tau_{\mathrm{diff}}$.
The dashed lines in Fig.\,\ref{SI_fig:4}\textit{A} fit $y = a \tanh \left[b \left(z^{\star} - c \right) \right] + d$ to the profiles (excluding the effect of the left-hand side boundaries) with $c$ the center of the buckling front and $w_{10-90} = 2.2/b$ its 10\%-to-90\% width; see right plot of Fig.\,\ref{SI_fig:4}\textit{A} for definitions.
In Fig.\,\ref{SI_fig:4}\textit{B} their numerical values are compared for the experiment (square) and simulation (circle). 
We find that the front width in the simulation is systematically smaller. 
An explanation for this might be the limited number of bonds in the 2D experiments.      
Apart from the center and the width, the slope between points in the left plot of Fig.\,\ref{SI_fig:4}\textit{B} contains information regarding the velocity of the buckling front, as $\epsilon = \epsilon_0 + \dot{\epsilon} t$.

Eventually, values of the non-dimensional in-plane diffusion coefficient ($\tilde{D}_{xz}$) and reaction term ($\tilde{\mathcal{R}}$) are determined that provide a match between experiments and simulations, where $D_{xz} / L_{\mathrm{bond}}^2 = \mathcal{O}(1)$ and $\mathcal{R} = \mathcal{O}(1)$. 

\section*{VI - Additional lattice configurations}
\begin{figure}[h!]
  \centerline{\includegraphics[width=\columnwidth]{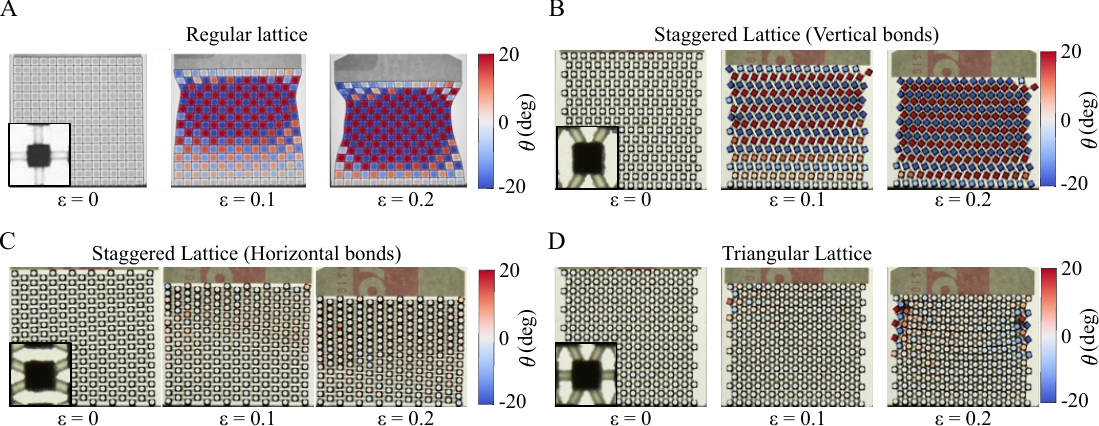}}
\caption{Lab-frame node rotation of 2D lattices similar to those in Fig.\,5 in the main text for a (\textit{A}) regular lattice, (\textit{B}) staggered lattice with vertical bonds, (\textit{C}) staggered lattice with horizontal bonds, and (\textit{D}) triangular lattice.
}
\label{SI_fig:5}
\end{figure}

To investigate the affect of lattice design on the formation and evolution of the diffusive buckling fronts in the experimental 2D systems, we have fabricated similar designs as shown in Fig.\,5 in the main text (here Fig.\,\ref{SI_fig:5}\textit{A}) but with slightly different configurations. We then perform the same compression experiments as discussed in the main text with a crosshead velocity of $v_{\mathrm{crosshead}} = \SI{0.5}{\mm \per \s}$ to a target strain of $\epsilon = 0.2$.
We then plot the lab-frame node rotations of all the nodes, overlayed on the experimental snapshot in Fig.\,\ref{SI_fig:5}.

In Fig.\,\ref{SI_fig:5}\textit{B} we consider a staggered lattice where the bonds are aligned vertically in the direction of the uni-axial compression.
Now, node rotations are still observed yet nodes within the same row co- instead of counter-rotate. 
Nonetheless, the counter-rotation of the nodes and the associated local buckling is still present, but now on the diagonals, parallel to the orientation of the bonds. 
Buckling still nucleates at the upper edges of the 2D lattice before propagating through the system, similar to the regular lattice, however the extend to which the lattice shrinks inwards during compression, e.g., the extend of its auxetic behaviour, reduces.

When rotating the staggered lattice the bonds align perpendicular the direction of the uni-axial compression, see Fig.\,\ref{SI_fig:5}\textit{C}. 
Now, no bonds are sufficiently load-bearing to induce buckling and node rotations remain absent throughout the compression.
In the continuum models this corresponds to $F_z$ (and hence $\mathcal{C}$) in Eq.\,4 of the main text to remain sufficiently small, that the reaction terms remain negative (sinks) indefinitely. 

Lastly, when allowing for a higher degree of connectivity in a triangular lattice (see Fig.\,\ref{SI_fig:5}\textit{D}), node rotations are largely suppressed as the nodes become too constrained.  
Rotation of the nodes only occurs at the lesser confined boundaries. 

\section*{VII - Inertial effects  }

In the main text we have neglected inertial effects given the small masses of the nodes and the low strain rates.
Here, we explore how the dynamics are affected if inertia was significant. 
We apply the same methodology as above for the case of uniaxial compression.
Apart from potential energy $U$ we assume an additional contribution that represents the kinetic energy of the nodes, \textit{i.e.}, $T = J_I\dot{\theta}^2/2$, with $J_I$ the moment of inertia.
Then, $L = T - U$ and we evaluate $\mathrm{d}/\mathrm{d}t \left(\partial_{\dot{\theta}_{i,j,k}} L\right) - \partial_{\theta_{i,j,k}} L  +  \partial_{\dot{\theta}_{i,j,k}} \mathcal{F} = 0 \rightarrow \mathrm{d}/\mathrm{d}t \left(\partial_{\dot{\theta}_{i,j,k}} T \right) + \partial_{\theta_{i,j,k}} U  +  \partial_{\dot{\theta}_{i,j,k}} \mathcal{F} = 0$ (equivalently for $\phi_{i,j,k}$).

Follow the same line of thought as above the final results read
\begin{equation*}
\begin{aligned}
J_I \ddot{\theta} + c_{\mathrm{disp}} \dot{\theta} = & \left(k_s L_{\mathrm{bond}}^2 - k_b \right)L_{\mathrm{bond}}^2 \nabla_{xz}^2 \theta + k_t L_{\mathrm{bond}}^2  \frac{\partial^2 \theta}{\partial y^2} + f(\theta,\phi), \\
J_I \ddot{\phi} + c_{\mathrm{disp}} \dot{\phi}=  & \left(k_s L_{\mathrm{bond}}^2 - k_b \right)L_{\mathrm{bond}}^2 \nabla_{yz}^2 \phi + k_t L_{\mathrm{bond}}^2   \frac{\partial^2 \phi}{\partial x^2} + g(\theta,\phi).
\end{aligned}
\end{equation*}
The forcing terms remain unchanged, apart from a prefactor. 
The above set of equations now resemble wave equations.
More specifically, they are non-linear Klein-Gordon equations with damping.
Equivalent forms for similar systems have been derived in earlier works \cite{deng2017elastic,deng2018metamaterials,mo2019cnoidal,deng2020characterization,deng2021dynamics,veenstra2024non}. 

\subsection*{Topological defects}

We assume buckling in only one angle and no damping to simplify the equation:
\begin{equation*}
J_I \ddot{\theta}  =  \left(k_s L_{\mathrm{bond}}^2 - k_b \right)L_{\mathrm{bond}}^2 \nabla_{xz}^2 \theta + k_t L_{\mathrm{bond}}^2  \frac{\partial^2 \theta}{\partial y^2} + N^2 k_c L_{\mathrm{bond}}^2 \epsilon \left[1 -  \frac{\epsilon_{\mathrm{crit}}}{\epsilon} - \frac{\theta^2}{6} \right]\theta.
\end{equation*}
Now we perform a coordinate substitution such that:
\begin{equation*}
\tilde{t} = \sqrt{\frac{\gamma}{J_I}}t, \quad \tilde{x} = \frac{\sqrt{\gamma}}{L_{\mathrm{bond}}\sqrt{k_s L_{\mathrm{bond}}^2 - k_b}}x, \quad \tilde{y} = \frac{\sqrt{\gamma}}{L_{\mathrm{bond}}\sqrt{k_t}}y, \quad \tilde{z} = \frac{\sqrt{\gamma}}{L_{\mathrm{bond}}\sqrt{k_s L_{\mathrm{bond}}^2 - k_b}}z, \quad \Theta = \frac{\theta}{\sqrt{-\frac{6 \gamma}{F L_{\mathrm{bond}}}}},
\end{equation*}
with $\gamma = N^2 k_c L_{\mathrm{bond}}^2\left(\epsilon - \epsilon_{\mathrm{crit}} \right)$.
Substitution will make the equation dimensionless, yielding
\begin{equation*}
\ddot{\Theta} = \nabla^2 \Theta + \Theta - \Theta^3.
\end{equation*}
Let us pick a line of propagation for our solution and assume planar symmetry around this line. 
Then our equation reads:
\begin{equation*}
\ddot{\Theta} = \partial^2_s  \Theta + \Theta - \Theta^3, \quad \text{where} \quad s = \begin{pmatrix} n_x \\ n_y \\ n_z \end{pmatrix} \cdot \begin{pmatrix} \tilde{x} \\ \tilde{y} \\ \tilde{z} \end{pmatrix}, \quad \text{and} \quad \begin{pmatrix} n_x \\ n_y \\ n_z \end{pmatrix} = \hat{n} \, \, \text{is a unit vector}. 
\end{equation*}
This equation has known solutions of the form \cite{lizunova2021introduction}
\begin{equation*}
\Theta(\tilde{t},s) = \pm \tanh \left( \frac{s - \tilde{a} + \tilde{v} \tilde{t}}{\sqrt{2\left(1 - \tilde{v}^2 \right)}} \right),
\end{equation*}
where $\tilde{a}$ and $\tilde{v}$ are dimensionless space and velocity variables.
Even for the static case, where $\tilde{v} = 0$, solutions are permitted and they have the form $\Theta = \pm \tanh \left(\left[s - \tilde{a}\right] / \sqrt{2} \right)$, which allow for the direction of buckling to be changed within certain planes.
It thus proves that the emergence of domain walls as seen in the experiments (see Fig.\,1\textit{D}) are an allowed static solution of our system of equations.

\subsection*{Dynamical stability analysis}

We would like to perform a stability analysis on the dynamical equation, considering only one angle to be non-zero and by neglecting cubic terms.
Therefore,
\begin{equation*}
J_I \ddot{\theta} + c_{\mathrm{disp}} \dot{\theta}  =  \left(k_s L_{\mathrm{bond}}^2 - k_b \right) L_{\mathrm{bond}}^2 \nabla_{xz}^2 \theta + k_t L_{\mathrm{bond}}^2  \frac{\partial^2 \theta}{\partial y^2} + N^2 k_c L_{\mathrm{bond}}^2 \left[\epsilon - \epsilon_{\mathrm{crit}} \right]\theta.
\end{equation*}
We can take the 4D Fourier transform to get:
\begin{equation*}
J_I \omega^2 + i c_{\mathrm{disp}} \omega = \left(k_s L_{\mathrm{bond}}^2 - k_b \right) L_{\mathrm{bond}}^2 \left(\kappa_x^2 + \kappa_z^2 \right) + k_t L_{\mathrm{bond}}^2 \kappa_y^2 - N^2 k_c L_{\mathrm{bond}}^2 \left[\epsilon - \epsilon_{\mathrm{crit}} \right],
\end{equation*}
with
\begin{equation*}
\omega = \frac{- i c_{\mathrm{disp}} \pm \sqrt{- c_{\mathrm{disp}}^2 + 4 J_I \left( \kappa_A^2 - m^2 \right)}}{2 J_I}, 
\end{equation*}
where $\kappa_A^2 = \left(k_s L_{\mathrm{bond}}^2 - k_b \right) L_{\mathrm{bond}}^2 \left(\kappa_x^2 + \kappa_z^2 \right)+ k_t L_{\mathrm{bond}}^2 \kappa_y^2$ and $m^2 = N^2 k_c L_{\mathrm{bond}}^2 \left[\epsilon - \epsilon_{\mathrm{crit}} \right]$.

Depending on the value of the determinant the system is underdamped, overdamped, or critically damped. This depends both on the material properties, such as the damping factor, the energy of the wave, but also on the amount of strain that is applied.

When $c_{\mathrm{disp}}^2 > 4 J_I \left(\kappa_A^2 - m^2 \right)$ the system is overdamped and the waves decay as exponentials.
If $c_{\mathrm{disp}}^2 = 4 J_I \left(\kappa_A^2 - m^2 \right)$ the system is critically damped and when $c_{\mathrm{disp}}^2 < 4 J_I \left(\kappa_A^2 - m^2 \right)$ it is underdamped and waves oscillate while losing energy.
Since $m^2 \sim \left[\epsilon - \epsilon_{\mathrm{crit}} \right] $ it can change sign depending on the amount of strain that is applied. 
During post-buckling, when $\epsilon > \epsilon_{\mathrm{crit}}$, the applied strain will cause the system to become more and more overdamped.

\end{document}